# Distributed Reactive Programming for Reactive Distributed Systems

## Florian Myter[a], Christophe Scholliers[b], and Wolfgang De Meuter[a]

a   Vrije Universiteit Brussel
b   Universiteit Gent

**Abstract**    **Context:** The term *reactivity* is popular in two areas of research: programming languages and distributed systems. On one hand, reactive programming is a paradigm which provides programmers with the means to declaratively write event-driven applications. On the other hand, reactive distributed systems handle client requests in a timely fashion regardless of load or failures.

**Inquiry:** Reactive programming languages and frameworks tailored to the implementation of distributed systems have previously been proposed. However, we argue that these approaches are ill fit to implement *reactive* distributed systems.

**Approach:** We analyse state of the art runtimes for distributed reactive programming and identify two key issues with regards to reactive distributed systems. They rely on single, central points of coordination and/or assume a lack of partial failures in the systems they support.

**Knowledge:** Based on our analysis we propose a novel runtime for distributed reactive programming languages and frameworks. This runtime supports reactive distributed systems by design.

**Grounding:** We implement a proof of concept framework for reactive distributed systems in JavaScript which builds atop our runtime. Using this framework we implement a case study application which highlights the applicability of our approach. Moreover, we benchmark our runtime against a similar approach in order to showcase its runtime properties and we prove its correctness.

**Importance:** This work aims to bridge the gap between two kinds of reactivity: reactive distributed systems and distributed reactive programming. Current distributed reactive programming approaches do not support reactive distributed systems. Our runtime is the first to bridge this reactivity gap: it allows for reactive distributed systems to be implemented using distributed reactive programming.

## The Art, Science, and Engineering of Programming

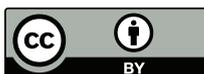

## 1 Introduction

The term *reactive* has been used to describe properties of both programming languages and systems. A programming language is reactive if it provides the constructs for programmers to declaratively implement event-driven applications. On the other hand, a system is reactive if it can respond to inputs in a timely fashion. For single-threaded non-distributed programs these definitions largely overlap. However, we observe that when moving to a distributed setting these definitions are different. We argue that the mismatch between these definitions has led to a situation where distributed reactive *programming languages* are ill fit to implement reactive distributed *systems*.

**Reactive Distributed systems**     Reactivity in terms of distributed systems relates to a set of design principles which aid programmers in developing responsive systems [5, 6, 16]. In other words, systems which respond to their clients' requests in a "timely fashion" regardless of load or partial failures in the system. Microservices [18] are an example of such systems. The following three requirements are crucial to the reactivity of a distributed system [5, 6, 16]:

**Resilience**  Failures are *isolated* to the concerned distributed component. This allows component failure to minimally impact the system as a whole. Client requests are still partially handled by the remaining components. Moreover, isolated failures facilitate the recovery of failed components.

**Elasticity**  The resources attributed to a particular component are dynamically increased or decreased depending on the load. *Decentralised* designs are favoured over centralised ones in order to replicate components as workload increases. This allows the system to handle client requests at the same pace regardless of variations in load.

**Asynchrony**  Communication between components happens through asynchronous messaging. This decouples the components in time and space, which facilitates concurrency and distribution respectively.

**Reactive Programming Languages**     Reactivity in terms of programming languages relates to the *reactive programming* (RP) paradigm [2]. In a nutshell, this paradigm allows programmers to elegantly write event-driven applications without resorting to callbacks or the observer-pattern (which are known to negatively impact code quality and program comprehension [28]). It does so by representing time-varying values (e.g. mouse position, system time, etc.) as first-class citizens called *signals*. Programmers declaratively combine signals using *signal combinators*, otherwise known as *lifted* functions. The language runtime tracks dependencies between signals in the form of a dependency graph. A propagation algorithm ensures that changes to a signal propagate through this graph, thereby updating the application. Using RP, one can implement each individual component of a distributed system (e.g. a single microservice).

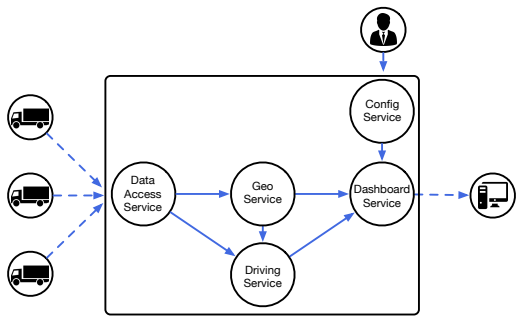

■ **Figure 1** Architecture of the fleet management application

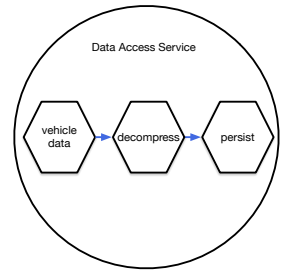

■ **Figure 2** The data access service's dependency graph

**The Distributed Reactivity Mismatch**   Distributed reactive programming (DRP) [8, 12, 27] extends the reactive programming paradigm to distributed systems: programmers declaratively specify dependencies between signals residing on distributed nodes in a network, which form a *distributed* dependency graph. In theory, this would allow programmers to declaratively implement the flow of data *across* components in a distributed system (e.g. how microservices coordinate with each other). However, current propagation algorithms for DRP are unfit to propagate changes in reactive distributed systems. Concretely, these algorithms use a central coordinator to propagate changes through the distributed dependency graph. They are therefore unable to meet the elasticity requirement (i.e. the centralised coordinator forms a bottleneck which negatively impacts scalability). Moreover, these algorithms are unable to handle partial failures in the distributed dependency graph. Instead, failures are total and affect the entire system. They are therefore also unable to meet the resilience requirement. To summarise: programmers cannot use current distributed reactive programming languages to implement reactive distributed systems.

This paper presents a novel propagation algorithm for DRP languages called QPROP. QPROP allows for the development of distributed reactive systems within the reactive programming paradigm. Such distributed systems cannot guarantee strong consistency. QPROP provides eventual consistency instead: when the system stops updating all nodes in the network eventually have a consistent view on the system's state. This paper provides the following contributions:

- The specification of a novel elastic, resilient and asynchronous propagation algorithm called QPROP.
- An implementation of a use case using a microservice framework in JavaScript build atop QPROP.
- A set of benchmark results, showing that QPROP outperforms state of the art DRP algorithms.
- A series of proofs demonstrating QPROP's correctness.

## 2   Current Reactive Languages and their Lack of Reactivity

This section discusses the shortcomings of current distributed reactive programming approaches. To this end we use a motivating example which stems from our cooperation with Emixis, a fleet management company, in the context of an industrial collaboration. Emixis' core business is to equip their customers' fleet of vehicles with tracking beacons and to provide software to analyse the data (e.g. GPS coordinates, current speed, etc.) uploaded by the beacons to a central server. The server is implemented using a microservice architecture, as shown in Figure 1. Concretely, these microservices have the following tasks:

**Data Access service**  Vehicles compress their sensory data before uploading it to the server. The *data access service* decompresses the uploaded data and subsequently persists the decompressed data.

**Geo service**  The vehicle data contains raw GPS coordinates. It is the *geo service*'s task to reverse geo code these coordinates into human-readable addresses.

**Driving service**  Parts of the vehicle data (e.g. speed, g-forces, etc.) are used to calculate eco-driving statistics by the *driving service*. Moreover, the service uses a vehicle's speed and its reverse geo-coded address in order to detect speed limit violations by the driver.

**Config Service**  Emixis' customers are able to configure certain aspects of their dashboard (e.g. icons used to represent vehicles, etc.). Customers specify their configurations by posting them to the *config* service.

**Dashboard service**  The reverse geo-coded addresses, driving statistics and customer configurations are combined by the *dashboard* service into a single view. This view is sent to the owner of the fleet of vehicles.

Using non-distributed reactive programming (e.g. [9, 21]) we can implement the individual microservices in our example. For instance, the *data access* service is implemented as follows. Vehicle data is represented as a signal, each time this signal changes the data is decompressed and persisted. We therefore use two lifted functions: one which decompresses the vehicle's data and one which persists the decompressed data. Figure 2 depicts the dependency graph constructed by the reactive runtime for the *data access* service. The runtime's propagation algorithm traverses the dependency graph in topological order as soon as a source signal changes. The algorithm updates each signal using its predecessors' values during this traversal. In our example, whenever the *vehicle data* signal changes, the algorithm updates the *decompress* signal with the new vehicle data. Subsequently, it updates the *persist* signal with the new decompressed data.

### 2.1  Distributed Propagation Algorithms and their Impact on Reactivity

Distributed reactive programming [8, 20, 27] applies the concepts of reactive programming to distributed systems. In other words, programmers are able to apply lifted functions on signals which reside on physically distributed machines. For example, using DRP the *geo* service applies a lifted reverse geo coding function on the *data* *access* service's *decompress* signal. Whenever the *decompress* signal changes (i.e. as a result of a change in a vehicle's data), the reverse geo coding function is invoked automatically (we say that the *geo* service *updates* whenever the *decompress* signal changes).

As explained in the previous section each microservice contains a dependency graph which represents its internal logic. However, using DRP the services themselves also form a *distributed* dependency graph. For example, Figure 1 shows the distributed dependency graph for our fleet management application. Whenever a vehicle's data changes, this updates the *data access* service which propagates the *decompressed* data to the *geo* and *driving* services. Subsequently, both services update before they propagate their new values further downstream to the *dashboard* service.

The topology of a dependency graph can dynamically change during the execution of a reactive program. These dynamic graph changes are used to implement a number of features in reactive programming languages and frameworks such as conditional propagation [9] and higher-order reactive values [19]. In the context of distributed reactive programming, a dynamic dependency graph is essential to allow components to connect to a running system (e.g. adding a microservice to our fleet management application while it is already online). We distinguish between two kinds of dynamic changes: involuntary temporal changes and intentional topological changes. The former occurs whenever a distributed component temporarily leaves the network (e.g. due to a disconnection). The latter occurs whenever a new distributed component joins the network or when an existing component chooses to leave the network.

As is the case for non-distributed dependency graphs, a propagation algorithm must perform a traversal in topological order of the distributed dependency graph to update all signals. To showcase the importance of this traversal order, consider the following hypothetical scenario. A vehicle sends its updated data to the *data access* service, which subsequently decompresses and persists this data. Assume that the *geo* and the *dashboard* services update before the *driving* service. The operator looking at the rendered dashboard might witness a faulty speed limit violation because the vehicle's position in the dashboard was updated before its driving statistics. This phenomenon is called a *glitch* [9]. A common strategy employed by non-distributed reactive programming languages to avoid glitches is to topologically sort the dependency graph before propagating values through it [9, 21].

However, this non-distributed strategy does not trivially scale towards reactive distributed systems. A single central coordinator would be required to explicitly sort the distributed dependency graph and to determine when each node may update. This approach is not resilient because the central coordinator constitutes a single point of failure in our system. Moreover, this central coordinator forms a performance bottleneck which goes against the elasticity requirement.

We distinguish two kinds of distributed reactive languages and frameworks: those which produce reactive but glitched distributed systems [8, 30] and those which produce glitch-free but unreactive distributed systems [3, 12]. An example of the former is AmbientTalk/R [8], which extends the actor-based AmbientTalk language with constructs for reactive programming. It targets peer-to-peer applications and is therefore decentralised by design. As a result it produces elastic systems, given that resources attributed to specific actors can grow and shrink. Moreover, actors can fail without hampering the system as a whole. Lastly, propagation of change in AmbientTalk/R happens asynchronously. Distributed systems implemented in AmbientTalk/R are reactive: they are elastic, resilient and asynchronous. However, AmbientTalk/R cannot guarantee glitch freedom and can therefore not support systems such as our fleet management application.

Distributed REScala and its SID-UP propagation algorithm [12] is the most prominent example of frameworks which produce glitch-free but unreactive distributed systems. First, SID-UP is unable to deal with failures of parts of the distributed network. As a consequence, failure of a single microservice would stop the propagation of change throughout the entire system. Systems implemented using distributed REScala are therefore not resilient. Second, SID-UP only guarantees glitch freedom if the source signals in an application do not update concurrently. If this is not the case a central entity is needed to coordinate access to the source signals. This coordinator only allows a value to propagate through the dependency graph once the previous value has *completely* traversed the graph. In other words, Distributed REScala can only support microservice systems where a single service handles all incoming requests sequentially. Systems implemented using distributed REScala are therefore not elastic. We discuss other approaches and frameworks in Section 8.

In this paper we present two propagation algorithms: QPROP and QPROP$^d$. Both algorithms ensure glitch freedom while meeting the requirements of reactive distributed systems. They differ in the kinds of dynamic graph changes they support. QPROP only supports temporal changes while QPROP$^d$ supports both temporal and topological changes.

Strongly consistent, non-reactive algorithms ensure that each update to a source signal results in all of the source's successors to update once. In contrast, QPROP and QPROP$^d$ are *eventually consistent* and allow for multiple concurrent updates to source signals to propagate through the distributed dependency graph. As a consequence, a signal in the dependency graph might only update once as a result of two source signals concurrently updating. However, once source signals stop updating QPROP and QPROP$^d$ guarantee the consistency of all signals in the distributed dependency graph.

We first discuss the implementation of our fleet management application before detailing QPROP in Section 4 and QPROP$^d$ in Section 5.

## 3   A Fully-Reactive Implementation of a Distributed Reactive System

To showcase the applicability of our approach we implement a part [1] of the fleet management application using Spiders.js [23]. This web-based actor framework written in TypeScript, a typed superset of JavaScript, provides built-in reactive programming constructs. It allows us to implement the internal logic of each individual service as well as the system's overall logic within the reactive programming paradigm.

---

[1] The entire implementation can be found at https://github.com/myter/ReactiveSpiders/tree/master (last accessed 2018-12-01).

### 3.1 Internal Service Reactivity

Listing 1 and Listing 2 provide the implementation of the *data access* service.

■ **Listing 1** Defining the fleet data signal

```
1  class FleetData extends Signal{
2    currentLat : number
3    currentLong : number
4    currentSpeed : number
5
6    @mutator
7    actualise(lat,long,speed){
8      this.currentLat = lat
9      this.currentLong = long
10     this.currentSpeed = speed
11   }
12 }
```

■ **Listing 2** The data access service's internal logic

```
1  class DataAccessService extends MicroService{
2    start(){
3      let dataSignal = this.newSignal(FleetData)
4      let decompressed this.fMap(decompress,
          ↪ dataSignal)
5      this.publish(decompressed)
6      this.fMap(persist,decompressed)
7      socket.on('message',(data) => {
8        dataSignal.actualise(data.lat,data.long,
            ↪ data.speed)
9      })
10   }
11 }
12 let monitor = new ServiceMonitor()
13 monitor.registerService(DataService)
14 //Install other services
15 monitor.deploy()
```

The service is implemented by extending the built-in *MicroService* class. All instances of *MicroService* are guaranteed to run under their own Node.js instance. A service's *start* method is called as soon as the service is instantiated. The *DataAccessService* accepts data packets sent by vehicles equipped with Emixis' beacons over UDP sockets. These packets are compressed by the vehicles, the service therefore decompresses them. Subsequently the decompressed data is persisted to avoid data loss. This cycle of decompressing and persisting needs to be performed each time a vehicle sends new data.

The implementation of the *DataAccessService* is structured according to the three main features of reactive programming. First, it represents time varying values as first class values called *signals*. In our example we represent the data sent by vehicles using instances of the *FleetData* signal class (see Listing 1 and line 3 in Listing 2). Second, the service derives new signals from existing ones using signal combinators: Spiders.js provides the *fMap* construct which has the following type signature:

$$(a \rightarrow b) \rightarrow Signal\ a \rightarrow Signal\ b$$

In other words, *fMap* lifts a function to be applied over signals. A lifted function is automatically re-evaluated as soon as one of its arguments changes. The *DataAccessService* lifts and applies two functions. The first function (i.e. *decompress*) decompresses the incoming data and is applied to the instance of *FleetData* (see line 4). The second function (i.e. *persist*) persists decompressed data, and is therefore applied to the signal which results from the previous lifted function application (see line 6). Lastly, changes to signals automatically propagate through the reactive application thereby re-evaluating lifted functions. Each time a vehicle sends new data the service triggers this propagation of change. It does so by invoking the data signal's annotated *actualise* method (see line 8). The *mutator* annotation informs Spiders.js that all invocations of the annotated method change the signal's state and should therefore trigger a propagation of change.

Spiders.js provides a service factory, called the *service monitor*, which instantiates services. This factory provides two main methods: *registerService* and *deploy*. The former registers a service class to be deployed (see line 13) while the later deploys all registered services (see line 15)

### 3.2 External Service Reactivity

In Spiders.js a lifted function can be applied over locally or remotely created signals (i.e. signals created by different microservices in the network). The semantics of lifted function application are the same regardless of the locality of its arguments: as soon as one of the arguments changes the lifted function is re-evaluated. Services publish and subscribe to signals using Spiders.js' topic-based publish-subscribe system. In a nutshell, each service is uniquely identified using a topic. Listing 3 contains the topic definitions for our example.

■ **Listing 3**  Defining the service topics

```
1  var DataTopic = new Topic("DataAccess")
2  var GeoTopic = new Topic("Geo")
3  var DrivingTopic = new Topic("Driving")
4  var ConfigTopic = new Topic("Config")
5  var DashboardTopic = new Topic("Dashboard")
```

These topics are used to invoke the service monitor's *registerService* method, which has the following type signature:.

$$Class \rightarrow topic \rightarrow [topic] \rightarrow nill$$

The method's parameters are the class definition of a service, its identifying topic and an array of topics to which the service subscribes. Listing 4 shows how this is applied to our example application.

■ **Listing 4**  Registering the services

```
1  let monitor = new ServiceMonitor()
2  monitor.registerService(DataAccessService,DataTopic,[])
3  monitor.registerService(GeoService,GeoTopic,[DataTopic])
4  monitor.registerService(DrivingService,DrivingTopic,[DataTopic,GeoTopic])
5  monitor.registerService(ConfigService,ConfigTopic,[])
6  monitor.registerService(DashboardService,DashboardTopic,[GeoTopic,DrivingTopic,ConfigTopic])
7  monitor.deploy()
```

The implementations of the individual services are oblivious to specific topics. Instead, services publish a signal using the *publish* method. On line 5 in Listing 2 the *data access* service publishes the *decompressed* signal. Moreover, references to subscribed signals are provided through a services' *start* method. As an example consider Listing 5, which provides the implementation of the *geo* service. Its start method provides a reference to the signal published by the *data access* service. This reference is used as argument to a lifted function which reverse geo codes coordinates (see line 3). Subsequently, the *address* signal is published and is subscribed to by the *driving* and *dashboard* services.

■ **Listing 5** Geo service implementation

```
1  class GeoService extends MicroService{
2    start(dataSignal){
3      let address = this.fMap(reverseGeoCode,dataSignal)
4      this.publish(address)
5    }
6  }
```

As our small example application showcases, distributed reactive programming allows programmers to coordinate multiple microservices. However, current DRP propagation algorithms are unable to uphold the reactivity requirements of modern microservice systems.

## 4 QPROP: A Propagation Algorithm for Reactive Distributed Systems

In this section we introduce a novel propagation algorithm for DRP called *queued propagation* (QPROP). QPROP is able to guarantee glitch freedom of distributed reactive applications while ensuring resilience, elasticity and asynchrony. Moreover, QPROP guarantees monotonicity and eventual consistency. However, QPROP does not guarantee progress and currently exhibits exponential computational complexity. We refer the reader to Appendix F for proofs and discussions regarding QPROP's guarantees.

Before detailing our algorithm we discuss key intuitions behind solving decentralised glitch freedom.

### 4.1 Decentralised Glitch Freedom: An Intuition

The key intuition behind glitches is that they can only occur for certain topologies of dependency graphs. Consider Figure 3(A), whenever $A$ propagates a value both $B$ and $C$ need to update before they propagate values to $D$. We say that $D$ is susceptible to glitches with regards to $A$. We refer the reader to Section 2.1 for a concrete example of this update ordering. In contrast, in Figure 3(B) no node is susceptible to glitches: $C$ can update as soon as it receives a new value from either $A$ or $B$ (which are both trivially free from glitches). For example, if $C$ receives a new propagation value from $A$ it uses the previously received value from $B$ to update itself.

In order to detect glitches, assume $A$ in Figure 3(A) possesses a logical clock which it increments each time it propagates a value to its successors. Moreover, $A$ tags each value it propagates with the current clock time. $U_D(val_B, val_C)$ denotes the updating

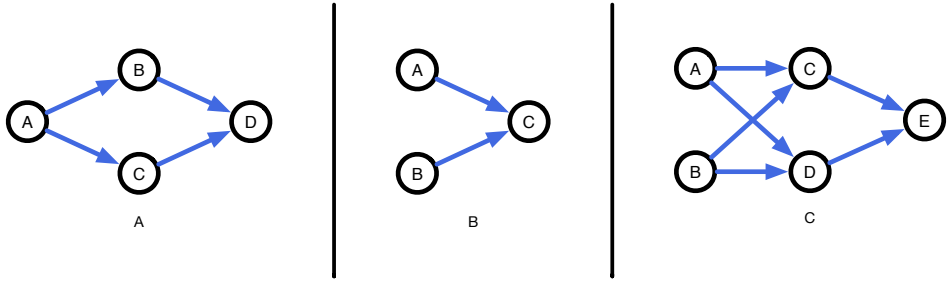

■ **Figure 3** A) A dependency graph susceptible to glitches. B) A dependency graph not susceptible to glitches. C) A dependency graph susceptible to concurrent glitches.

of $D$ using a value $val_B$ propagated by $B$ and a value $val_C$ propagated by $C$. $D$ must adhere to the following constraint to avoid glitches, where $Time_A(v)$ denotes $A$'s clock time tagged to value $v : U_D(val_B, val_C) \iff Time_A(val_B) == Time_A(val_C)$. Furthermore, values can propagate concurrently through the distributed dependency graph. Consider Figure 3(C), $A$ and $B$ might concurrently propagate new values. Given network delays these values can be received by $C$, $D$ and $E$ at arbitrarily different points in time. $E$'s constraint to avoid glitches is therefore the following:

$$U_E(val_C, val_D) \iff Time_A(val_C) == Time_A(val_D) \wedge Time_B(val_C) == Time_B(val_D)$$

To summarise, the concurrent and asynchronous nature of distributed systems can lead to glitches which would not occur in non-distributed systems.

### 4.2 QPROP

QPROP is divided into three phases. First, during the *exploration* phase each node uses its neighbours to explore its position in the graph. Second, the *barrier* phase ensures that the exploration phase has successfully finished before nodes start to propagate values. Third, the *propagation* phase ensures the actual propagation of values. QPROP assumes the following:

- As is commonly the case in reactive programming, dependency graphs are finite and acyclic [11, 12, 21]. Moreover, there are no intentional topological changes to the graph after the exploration phase.
- Nodes in the dependency graph are the unit of distribution (e.g. a microservice, a process running on a server, etc.).
- Propagation of values within a single node (i.e. through a non-distributed dependency graph) is abstracted as an update function. We assume that this update function provides glitch free propagation of values *within* nodes.
- Nodes communicate with one another through an asynchronous communication medium which ensures exactly once, in-order delivery of messages.
- At the start of the application (i.e. before the exploration phase) each node has references to its direct predecessors and successors in the graph. References uniquely identify nodes (e.g. references are IP addresses).

#### 4.2.1 Notation

We represent a node $n$ as a 9-tuple: $n = (DP, DS, I, S, U, initVal, lastProp, self, clock)$. Each element in the tuple contains the following information:

$DP$ is the set of $n$'s direct predecessors.

$DS$ is the set of $n$'s direct successors.

$I$ is a dictionary of input sets which stores values propagated by $n$'s direct predecessors (i.e. $I$'s keys are references to predecessors).

$S$ is a dictionary where the keys are references to source nodes and the values are sets of references to direct predecessor which are included in paths from the key source node to $n$.

$U$ is $n$'s update function, its arity equals $DP$'s cardinality. Once called with the values of $n$'s predecessors this function returns a single value to be propagated downstream by $n$.

$initVal$ is $n$'s initial value (i.e. the value before its first update).

$lastProp$ is $n$'s last propagated value.

$self$ is a reference to $n$.

$clock$ is a logical clock which $n$ uses to timestamp propagation values.

Table 1 in Appendix A provides a summary of these elements, we advise readers to keep it at hand while reading the rest of this section.

We describe our algorithm using pseudocode notation. Each node in the distributed dependency graph runs the phases of the algorithm in sequence (i.e. first the exploration phase, then the barrier phase and finally the propagation phase). During these phases nodes coordinate through asynchronous messaging. We provide pseudocode notation of message handlers to describe a node's behaviour upon receiving a particular message. We denote sending an asynchronous message $m$ with arguments $(arg_1, ..., arg_n)$ to a node $n$ using: $n \leftarrow m(arg_1, ..., arg_n)$. Moreover, using $await$ we specify that the execution of the pseudocode only continues once $n$'s message handler returns (e.g. $value = await\ n \leftarrow m(arg_1, ..., arg_n)$). To denote the elements within a dictionary we use the following notation: $[k, v] \in D$ where $[k, v]$ is a key-value pair and $D$ is a dictionary. To read the value bound to key $k$ in dictionary $D$ we write: $D.k$. Finally, to add a key-value pair $[k, v]$ to a dictionary $D$ we write: $D = D \cup \{[k, v]\}$.

### 4.2.2 Exploration Phase

---

**ALGORITHM 1:** Exploration

1  *sourcesReceived = 0*
2  **foreach** *pred* ∈ *DP* **do**
3    │  *I = I ∪ {[pred, {}]}*
4  **end**
5  **if** *|DP| == 0* **then**    /* I am a source node */
6    *lastProp = (self, initVal, {[self], 0}, 0)*
7    **for** *succ* ∈ *DS* **do**
8      │  *succ ← sources({self}, lastProp)*
9    **end**
10 **end**

---

**Handler** sources(sources,initProp)

1  *from = initProp.from*
2  *I.from = I.from ∪ {initProp}*
3  *sourcesReceived += 1*
4  **foreach** *s* ∈ *sources* **do**
5    **if** *[s, _] ∉ S* **then**
6      │  *S = S ∪ {[s, {}]}*
7    **end**
8    *S.s = S.s ∪ {from}*
9  **end**
10 **if** *sourcesReceived == |DP|* **then**
11    *allSources = {s|[s, _] ∈ S}*
12    *sourceClocks = {[s, 0]|s ∈ allSources}*
13    *lastProp = (self, initVal, sourceClocks, 0)*
14    **for** *succ* ∈ *DS* **do**
15      │  *succ ← sources(allSources, lastProp)*
16    **end**
17 **end**

---

Algorithm 1 provides the specification of a node's behaviour during the exploration phase. The algorithm is executed for each node at the start of the reactive program. Only the node's direct predecessors, direct successors, initial value, self reference and logical clock are known at this point (i.e. *DP*, *DS*, *initVal*, *self* and *clock* contain this information, all other node elements are empty).

Informally the purpose of the algorithm is twofold. First, each node computes the paths from source nodes which lead to that node. Second, nodes populate their *I* dictionaries with their predecessors' initial values.

At the start of the exploration phase each node creates a new dictionary per predecessor to store that predecessor's input set (see line 2 to line 4). Moreover, source nodes send the *sources* message (see line 5 to line 9) which contains a singleton set with their self reference and their initial propagation value. We represent propagation values as 4-tuples *propVal = (from, value, sClocks, fClock)*: *from* is a reference to the node propagating the *value*, *sClocks* is a dictionary of clock times for all source nodes which are direct or indirect predecessors of *from* and *fClock* is *from*'s clock time.

The *sources* Handler defines how nodes handle the *sources* message. As soon as a node receives a set of source references from each of its direct predecessors it relays these references together with its initial propagation value to all its direct successors (see line 10 to line 16 in the *sources* Handler). At the end of this process each node in the graph knows which source nodes are able to reach it and through which direct predecessor. For example, in Figure 3(C) *E*'s *S* dictionary contains $[A, \{C, D\}]$ and $[B, \{C, D\}]$

### 4.2.3 Barrier Phase

---

**ALGORITHM 2:** Barrier

1  *startsReceived = 0*
2  **if** *|DS| == 0 ∧ sourcesReceived == |DP|* **then**
3    **foreach** *pred* ∈ *DP* **do**
4      │  *pred ← start()*
5    **end**
6  **end**

---

**Handler** start()

1  *startsReceived += 1*
2  **if** *startsReceived == |DS|* **then**
3    **foreach** *pred* ∈ *DP* **do**
4      │  *pred ← start()*
5    **end**
6  **end**

---

Glitches could occur if values were to propagate before all nodes were able to construct their input queues. The barrier phase, see Algorithm 2, allows source nodes to determine when it is safe to start propagating values.

As soon as a sink node (i.e. a node without successors) is done exploring (i.e. it has received a *sources* message from all its direct predecessors, see line 2), it sends a *start* message to all its predecessors to indicate that they can start producing values. Any non-sink node relays this message upstream as soon as it has received a *start* message from *all* its direct successors (see line 2 in the *start* Handler). A source node starts propagating values once it receives a *start* message from all its direct successors. At this point the source node knows by induction that all downstream successors are done exploring.

### 4.2.4 Propagation Phase

---
**Handler** change($v_{new} = (from, value, sClocks, fClock)$)

---
1   $I.from = I.from \cup \{v_{new}\}$
2   $allArgs = \times(\{\{v_{new}\}\} \cup \{i | i \in I \setminus \{I.from\}\})$
3   $matches = \{args \in allArgs | \forall arg_{dp_1}, arg_{dp_2} \in args : [s, \{dp_1, ..., dp_2\}] \in S \rightarrow arg_{dp_1}.sClocks.s == arg_{dp_2}.sClocks.s\}$
4   **if** $matches \neq \emptyset$ **then**
5      $lastMatch = max(matches)$
6      $clock += 1$
7      $lastProp = (self, U(lastMatch.values), lastMatch.sClocks, clock)$
8      **foreach** $succ \in DS$ **do**
9         $succ \leftarrow change(lastProp)$
10      **end**
11      **foreach** $arg = (f, v, sc, fc) \in lastMatch$ **do**
12         $I.f = I.f \setminus \{vals \in I.f | vals.fClock < fc\}$
13      **end**
14 **end**

---

Each key-value pair $[s, preds] \in S$ informs a node that it can only update using values received from predecessors in *pred* if these values have equal clock times for source node $s$. In other words, the following must hold:

$U(arg_{pred_1}, ..., arg_{pred_n}) \iff \forall [s, \{pred_i, ..., pred_j\}] \in S : arg_{pred_i}.sClocks.s == arg_{pred_j}.sClocks.s$

The *change* Handler defines the heart of QPROP (i.e. how values propagate through the distributed dependency graph in a glitch free way). It starts as soon as a node $n$ receives a new propagation value $v_{new} = (from, value, sClocks, fClock)$ which is immediately stored in $n$'s $I$ set for *from* (see line 1). We assume that each input set in $I$ is totally ordered based on its values' *fClocks*. Subsequently, $n$ computes a nested set of all possible arguments to its update function (see line 2). This partially ordered set is obtained by taking the cross product (marked by the × operator on line 2) of $v_{new}$ with each of $n$'s predecessors' $I$ sets.

$n$ filters this set of arguments to only contain glitch-free sets of arguments (see line 3). Furthermore, $n$ takes the lexicographic maximum of these sets of arguments (see line 5). This set contains the last value propagated by each predecessor which can be used to update $n$ in a glitch free way. $n$ uses these values to invoke its update function after which it propagates its updated value to all direct successors (see line 7 to line 10). We assume that *lastMatch.values* and *lastMatch.sClocks* respectively return a set containing the value of each argument in *lastMatch* and the union of the *sClocks* of each argument in *lastMatch*.

Finally, *n* removes all stale values from its input sets. We consider a value to be stale if a newer value propagated by the same predecessor has previously been used as argument to *U*. All values older than the ones in *lastMatch* are stale and can therefore be removed (see line 11). We refer the reader to Appendix B for a concrete example of QPROP's workings.

In conclusion, QPROP meets the reactivity requirements as follows:

**Resilience**  The temporary failure of a node in the distributed dependency graph does not cause a system-wide failure. Instead, only the failed node's downstream successors might stop propagating values (i.e. if the failed node's values are required to avoid glitches). For example, if the *config* node in our fleet management application fails the *dashboard* node will continue updating as the vehicles' data changes. Although QPROP does not explicitly provide fault-tolerance mechanisms, it can be implemented atop existing crash-recovery frameworks [1, 15, 31].

**Elasticity**  In QPROP nodes are completely autonomous. Each node propagates values at its own pace without intervention from a central coordinator, provided that this propagation does not cause a glitch. For example, if the *geo* node in our fleet management application slows down because of request load one can dynamically increase its computational resources to scale the application.

**Asynchrony**  QPROP exclusively relies on asynchronous communication between nodes.

## 5  Supporting Dynamic Graph Changes with QPROP$^d$

QPROP only supports involuntary temporal changes. To support intentional topological changes we extend QPROP with QPROP$^d$ (i.e. dynamic QPROP). We support four dynamic operations: adding a node to and removing a node from the dependency graph and adding a dependency to and removing a dependency from the graph. For the sake of brevity we only discuss the dynamic addition of a dependency in this section, all other operations are provided in Appendix D.

### 5.1  Dynamic Graph Changes: An Intuition

Applying dependency changes to the graph's topology requires all directly or indirectly affected nodes to update their *S* and *I* dictionaries. Figure 4(A) and (B) respectively show the state of the dependency graph before and after the addition of a new dependency between *A* and *D*. It is QPROP$^d$'s task to extend *E*'s *S* dictionary entry for *A* from [*A*, {*C*}] to [*A*, {*C*,*D*}], to add *A* to *D*'s *S* dictionary and to ensure that *E* now receives values originating from *A* through *D*.

Note that the system is running, QPROP$^d$ must therefore guarantee glitch freedom during the addition of dependencies. Assume that source node *A* propagates value *a₁* before the addition of the dependency (as shown in Figure 4(A)). Throughout this example we omit the *sClocks* and *fClock* parts of propagation values for the sake

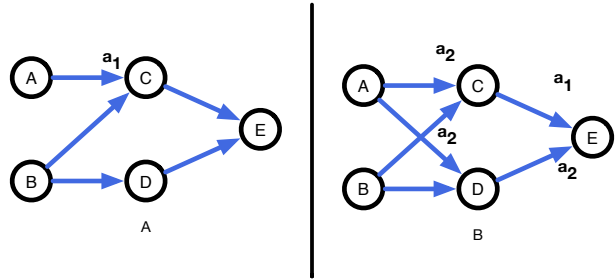

■ **Figure 4** A) Dependency graph before dynamic addition of a dependency between *A* and *D*. Source *A* propagates value $a_1$. B) Dependency graph after dynamic addition of a dependency between *A* and *D*. Source *A* propagates value $a_2$.

of brevity. We also assume that *B* does not change throughout the example (e.g. *C* updates itself using $a_1$ and *B*'s initial value as arguments). Due to network congestion, node *E* is yet to receive $a_1$ from *C*.

*A* propagates value $a_2$ after the dependency between *A* and *D* is added and all *S* dictionaries have been updated. $a_1$ finally arrives to *E* via *C* and $a_2$ arrives to *E* via *D*. However, dynamically adding a dependency changed *E*'s *S* dictionary. As a result, *E* can now only update if:

$$\exists arg_C \in I.C, \exists arg_D \in I.D : arg_C.sClocks.A == arg_D.sClocks.A.$$

*D* will never propagate $a_1$, given that it was not a successor of *A* at the time. This leaves *E* with two options. First, *E* continues behaving according to the definition of QPROP. In this case *E* can only satisfy the aforementioned condition using $a_2$ values. *E* will never update using *C*'s $a_1$ value and will remove it from its *I* sets instead (see the *change* Handler). In other words, the dynamic addition of a dependency resulted in the loss of a message.

A second option is for *E* to temporarily ignore values propagated by *D*. As soon as *E* has updated itself with *C*'s $a_1$ and *D*'s last valid propagated value (i.e. not *D*'s $a_2$ value) it can resume regular propagation using values from *C* and *D*. *E* can safely use this value for *D* given that it was propagated before the addition of the dependency between *A* and *D*. We say that *D* is *brittle* for *E*. Concretely, the problem occurs when a node must extend an already existing entry in its *S* dictionary. In our example, node *E* adds *D* to the $[A, \{C\}]$ entry in *S*. To avoid the loss of messages we choose for this latter option in QPROP[d].

## 5.2 QPROP[d]

---

**ALGORITHM 3:** Dynamic dependency addition

---
  **arguments :** A new predecessor *pred*
1. $(predLastProp, sources) =$ **await** $pred \leftarrow newSucc(self)$
2. $DP = DP \cup \{pred\}$
3. **if** $\nexists source \in sources : [source, \_] \in S$ **then**
4.   |    $I = I \cup \{[pred, \{predLastProp\}]\}$
5. **end**
6. **await** $self \leftarrow addSources(pred, sources)$

---

**Handler** newSucc(succ)

---
1. $DS = DS \cup \{succ\}$
2. **if** $|DP| == 0$ **then**
3.   |    **return** $(lastProp, \{self\})$
4. **else**
5.   |    $allSources = \{s | [s, \_] \in S\}$
6.   |    **return** $(lastProp, allSources)$
7. **end**

---

**Handler** addSources(from,sources)

---
1. **foreach** $source \in sources$ **do**
2.   |    **if** $[source, \_] \in S$ **then**
3.   |   |    $S.source = S.source \cup \{from\}$
4.   |   |    $B_r = B_r \cup \{[from, \{\}]\}$
5.   |    **else**
6.   |   |    $S = S \cup \{[source, \{from\}]\}$
7.   |    **end**
8. **end**
9. **foreach** $succ \in DS$ **do**
10.   |    **await** $succ \leftarrow addSources(self, sources)$
11. **end**

---

For each dynamic operation in QPROP[d] we provide an algorithm which is run by the node initiating the operation (e.g. a node dynamically adding a dependency to a new predecessor). Moreover, we define a number of new message handlers which extend the set of message handlers defined in Section 4.

A node $n$ dynamically adding a dependency to a new predecessor *pred* runs Algorithm 3, which performs three main tasks. First, $n$ requests the last propagated value from *pred* together with a set of all sources able to reach *pred* (see line 1). By requesting this information, *pred* adds $n$ to its list of successors (see line 1 in the newSucc Handler). Second, if $n$ is not brittle for *pred* (i.e. there is no overlap between the sources that can reach *pred* and those that can reach $n$) it creates an entry in $I$ for *pred* (see line 3). Third, $n$ updates its own topological information (i.e. the $S$ dictionary) and that of its direct and downstream successors. To do so , $n$ sends itself the *addSources* message (see line 6) with *pred* and the sources which can reach *pred* as arguments. The addSources Handler uses these arguments to update the receiving node's $S$ dictionary (see line 2 to line 6). Moreover, the handler tracks which predecessors are brittle for $n$ in a dictionary $B_r$ which contains entries $[brittlePred, vals] \in B_r$, where *brittlePred* is a brittle predecessor for $n$ and *vals* are values propagated by said predecessor. This *addSources* message is recursively sent to $n$'s direct and downstream successors (see line 10).

$n$ immediately continues with QPROP[d]'s pre-propagation phase once the dependency addition operation has completed (i.e. $n$ and all its downstream successors have updated their topological information). A barrier phase is not required here, given that values are already propagating through the system.

QPROP[d] introduces a pre-propagation phase for all nodes. The goal of this phase is to determine whether certain predecessors cease to be brittle as the result of a node receiving a new propagation value. Consider the example depicted in Figure 4(B). $D$ ceases to be brittle for $E$ as soon as $E$ updates using $C$'s $a_1$ value as arguments. In other words, when the value with the smallest clock time for $A$ in $B_r.D$ is one clock

---

**ALGORITHM 4:** Pre-propagation

---

**arguments :** A new value $v_{new} = (from, value, sClocks, fClock)$

1  **case** $\neg isBrittle(from) \wedge \neg hasBrittleSibling(from)$ **do**
2     |  $self \leftarrow change(v_{new})$ /* Proceed with QPROP's propagation phase */
3  **case** $\neg isBrittle(from) \wedge hasBrittleSibling(from)$ **do**
4     |  $I.from = I.from \cup \{v_{new}\}$
5     |  **if** $\forall pred \in DP : isBrittleSibling(pred, from) \implies B_r.pred \neq \emptyset$ **then**
6     |  |  $self \leftarrow change(v_{new})$ /* Proceed with QPROP's propagation phase */
7     |  |  **foreach** $pred \in DP : isBrittleSibling(pred, from) \wedge synchronised(pred)$ **do**
8     |  |  |  $MoveToI(pred)$
9     |  |  **end**
10    |  **end**
11 **case** $isBrittle(from)$ **do**
12    |  $B_r.from = B_r.from \cup \{v_{new}\}$
13    |  **if** $|B_r.from| == 1$ **then**
14    |  |  **if** $synchronised(from)$ **then**
15    |  |  |  $MoveToI(from)$
16    |  |  |  $self \leftarrow change(v_{new})$ /* Proceed with QPROP's propagation phase */
17    |  |  **else**
18    |  |  |  **foreach** $pred \in DP : isBrittleSibling(from, pred)$ **do**
19    |  |  |  |  **foreach** $val \in I.pred \setminus \{I.pred.first()\}$ **do**
20    |  |  |  |  |  /* Run Algorithm4 with val as input */
21    |  |  |  |  **end**
22    |  |  |  **end**
23    |  |  **end**
24    |  **end**
25 **end**

---

time bigger than the value with the smallest clock time for $A$ in $I.C$ (we assume $a_1$ and $a_2$ to have clock times for $A$ of 1 and 2 respectively). We say that $D$ is a brittle sibling of $C$ and that $D$ has synchronised with $C$ if the aforementioned condition holds.

Algorithm 4 defines a node $n$'s behaviour when it receives a *change* message from a predecessor *from*. It relies on a number of predicates (i.e. *isBrittle, hasBrittleSibling, isBrittleSibling* and *synchronised*) and a function (i.e. *MoveToI*) which are defined in Appendix D.1. $n$ executes the algorithm before the *change* Handler. The algorithm discriminates based on the predecessor *from* propagating a value $v_{new}$ to $n$:

1. *from* is not a brittle predecessor and it does not have any brittle siblings (see line 1). In this case $n$ can safely continue with QPROP's propagation phase.

2. *from* is not a brittle predecessor but it has at least one brittle sibling (see line 3). In other words, $n$ has another predecessor *pred* which is brittle and which shares a common predecessor with *from*. The algorithm first checks whether all brittle siblings of *from* have at least propagated one value. Assume a brittle predecessor $pred_{brittle}$ has not yet propagated a value (i.e. $B_r.pred_{brittle}$ is empty). It is impossible for $n$ to assess whether $pred_{brittle}$ is already synchronised or whether it is still brittle and $n$ can therefore not update itself safely. If all brittle siblings have at least propagated one value, $n$ is able to try and update itself with $v_{new}$. Subsequently the algorithm checks whether certain predecessors have ceased to be brittle as a result of this possible update. If a predecessor has ceased to be brittle (indicated by the *synchronised* predicate) the *MoveToI* function is invoked which copies the predecessor's propagation values from $B_r$ to $I$ and removes it from $B_r$.

3. *from* is a brittle predecessor (see line 11). The algorithm starts by checking whether $v_{new}$ is the first message received from *from* (i.e. $|B_r.from| == 1$). If this is the case it can be that *from* is already synchronised (see line 14) in which case the *MoveToI* function is invoked and $n$ is able to safely update. This would be the case in our example if $D$ would propagate $a_1$ due to $A$ only propagating $a_1$ after the dynamic dependency addition. If *from* is not synchronised but $v_{new}$ is the first message received from *from* (see line 18) the algorithm needs ensure that *from*'s non-brittle siblings have no unprocessed values. Concretely, when $B_r.from$ is empty all values propagated by its non-brittle siblings are stored in $I$ without being processed (see line 4). Therefore, when $n$ receives the first value from *from* the algorithm needs to process these unused values.

In summary, QPROP$^d$ relaxes one of the assumptions made by QPROP. Namely, that the dependency graph does not intentionally change once the system has passed the exploration phase. However, this relaxation comes at the price of temporarily ignoring propagation values. This is a necessary evil in order to guarantee glitch freedom in this dynamic context.

## 6 Evaluation

Our evaluation of QPROP and QPROP$^d$ is twofold. First, we prove that QPROP and QPROP$^d$ guarantee glitch freedom, eventual consistency, monotonicity and absence of progress (see Appendix F). Second, we compare the runtime performance of our approach to that of SID-UP [12]: a state of the art centralised approach. As we discuss in Section 2, approaches such as the one proposed in [3] essentially rely on the same central coordination technique. Moreover, as shown in [12] the algorithm outperforms adaptations of non-distributed propagation algorithms.

In order to compare our approach to SID-UP we systematically compare the runtime performance of two distributed reactive systems implemented in Spiders.js: one built atop QPROP or QPROP$^d$ and one built atop SID-UP. We first compare a QPROP and SID-UP implementation of the fleet management application detailed Section 3. Subsequently we compare a larger, artificial, application built atop QPROP, QPROP$^d$ and SID-UP. We compare the approaches using the following three metrics:

**Load** is the amount of requests per second the system receives. Each request results in the propagation of a value through the distributed dependency graph.

**Latency** is the average time it takes for a single value to propagate from a given source node to a sink node.

**Throughput** is the amount of values which propagate from source node to sink node for a given period of time.

**Processing time** is the time it takes for a request to propagate to a sink node. The difference with latency is that we start measuring processing time as soon as a request has been made. In contrast, we measure latency only as soon as the request is first propagated by a source node.

**Memory usage** we define the *heap memory usage* as the memory used by a particular service in its allocated heap. Moreover, we define the *RSS (Resident Set Size) memory usage* as the memory used by a particular service in its allocated heap, stack and code segment. Both heap memory usage and RSS memory usage are measured through node.js' *process.memoryUsage()* [2]

It is important to note that due to QPROP and QPROP$^d$'s eventually consistent nature they might perform less updates than SID-UP for a same load. In SID-UP each update of a source node is guaranteed to cause a single update for all its successors. In QPROP and QPROP$^d$ concurrent updates to multiple source nodes might only cause common successors to update once. To ensure the fairness of our comparison we therefore only consider a benchmark to have finished when the given system has completely processed the given amount of load. For example, assume a benchmark which simulates a load of 100 requests per second for a total of 30 seconds. Moreover, the system used for the benchmark contains a single sink node. For both SID-UP as well as QPROP and QPROP$^d$ we consider the benchmark to have finished if the system's sink node has updated 3000 times. In Appendix E.3 we measure the amount of these *concurrent interactions*.

In general our benchmarks show that QPROP significantly outperforms SID-UP with regards to throughput, processing time and memory usage. This is in spite of QPROP's substantial computational complexity which is $\mathcal{O}(M^N)$ for QPROP's core (i.e. line 2 in the *change* Handler) where $M$ is the worst-case amount of messages stored for $N$ direct predecessors. Although it slightly under performs with regards to latency, QPROP is able to respond to client requests in a timely fashion regardless of load. As a result, a distributed system implemented atop QPROP is more reactive than its SID-UP variant.

### 6.1 Use Case Comparison

The goal of our first benchmark is to compare both approaches in a real-world setting. We therefore compare runtime performance for our fleet management system presented in Section 3, which is a prototypical implementation of the production system deployed by Emixis. We measure latency, throughput, processing time and memory usage under varying loads actually measured by Emixis' production version of the fleet management application. In a nutshell the production system receives on average 45 requests per second during the weekend, 75 requests per second during the evening and 300 requests per second during daytime. We conduct the benchmarks using a setup similar to Emixis', namely an Ubuntu 14.04 server with two dual core Intel Xeon 2637 processors (2 physical threads per core) with 265 GB of RAM.

Figure 5 and Figure 6 respectively show how both systems compare with regards to throughput and latency under varying load. The former clearly shows that QPROP outperforms SID-UP with regards to throughput. SID-UP is unable to handle more than

---

[2] https://nodejs.org/api/process.html#process_process_memoryusage (last accessed 2018-12-01).

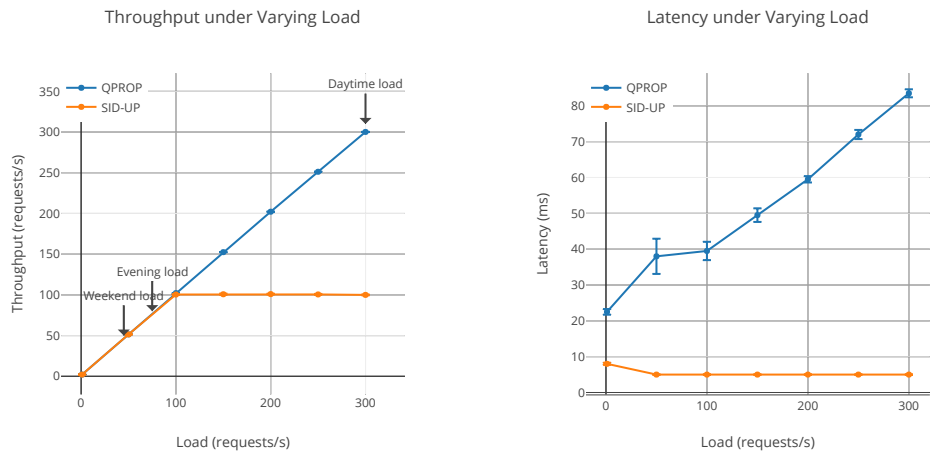

■ **Figure 5** Throughput under varying loads. Error bars indicate the 95 % confidence interval

■ **Figure 6** Latency under varying loads. Error bars indicate the 95 % confidence interval

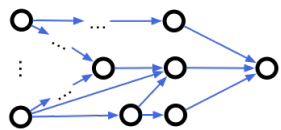

■ **Figure 7** Dependency graph of the larger microservice system

100 requests per second, all additional requests remain in the central coordinator's buffer. In other words, SID-UP is unable to efficiently handle the daytime load of our fleet management application. However, QPROP performs worse with regards to latency. In SID-UP a value can only propagate through the distributed dependency graph when the previous value has completed its propagation. In other words, all nodes in the distributed dependency graph are always ready to accept new values. Latency is therefore unaffected by the load. In QPROP this is not the case, values propagate through the graph concurrently. Upon receiving a new value, a node could still be processing the previous one which negatively impacts latency. Additional benchmark results are provided in Appendix E.1 (QPROP also outperforms SID-UP with regards to processing time and memory usage).

## 6.2 General Comparison

The fleet management application only consists of five microservices. To further investigate the performance properties of QPROP, QPROP[d] and SID-UP we compare the approaches using a larger example. Concretely, we implement a system comprised of 60 microservices where each service relays the requests it receives to other services. The application's dependency graph is exemplified in Figure 7. We conduct the benchmarks on a cluster of 60 Raspberry Pi 3 devices (Quad Core 1.2 GHz Broadcom 64 bit CPU

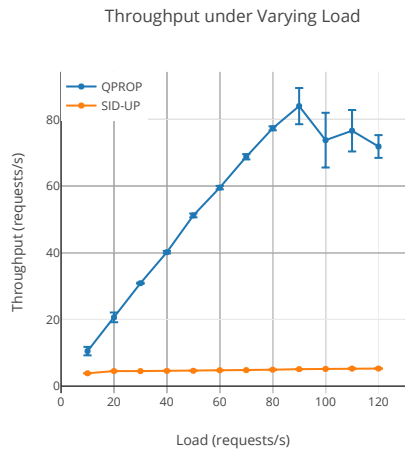

Throughput under Varying Load

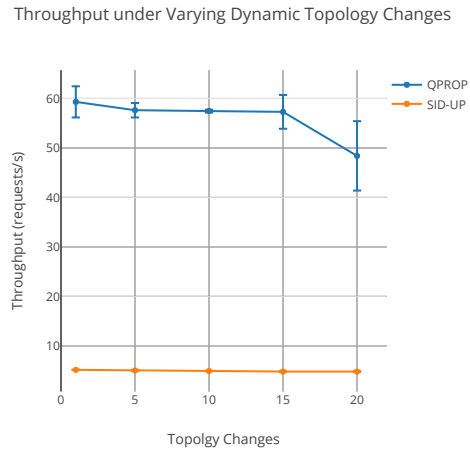

Throughput under Varying Dynamic Topology Changes

■ **Figure 8** Throughput under varying loads. Error bars indicate the 95 % confidence interval

■ **Figure 9** Throughput under varying dynamic topology changes. Error bars indicate the 95 % confidence interval and 1 GB of RAM). Each Raspberry Pi has a 100 Mbit/s network port and hosts a single microservice.

Figure 8 compares the results of the QPROP and SID-UP implementations of the microservice systems. As is the case for the fleet management application, SID-UP is able to handle considerably less load than QPROP. SID-UP's maximum throughput is roughly 5 requests per second, while QPROP reaches its throughput peak at 85 requests per second. Although QPROP's throughput decreases after this peak, it is still roughly able to handle an order of magnitude more requests per second.

In order to compare QPROP[d] and SID-UP we measure the impact of performing operations which intentionally change the dependency graph's topology. For a static load of 100 requests per second we vary the number of dynamic dependencies added to the dependency graph. Figure 9 shows the results of these experiments. Dynamic topology changes affect both systems' throughput: QPROP's throughput roughly decreases with 20 % when 20 operations are performed while SID-UP's throughput decreases with roughly 10 %. Additional results for these benchmarks are discussed in Appendix E.2.

## 7 Limitations

QPROP guarantees eventual consistency (see Appendix F for a proof). Assuming that source nodes stop propagating new values, all nodes in the graph eventually update using their predecessors' last values. However, nodes in QPROP can become subject to livelocks. For example, reconsider Figure 3(C). Node $E$'s glitch freedom constraint goes as follows:

$$U_E(val_C, val_D) \iff Time_A(val_C) == Time_A(val_D) \wedge Time_B(val_C) == Time_B(val_D)$$

However, it is possible that $E$ never receives a pair $(val_C, val_D)$ for which this holds. Assume that $A$ and $B$ update concurrently. Due to interleaving of messages it can be that $C$ invokes its update function using the new value for $A$ and $B$'s old value as arguments. Furthermore, $D$ invokes its update function using the new value for $B$ and the old value for $A$ as arguments. Upon receiving these values, $E$ will not be able to meet its glitch freedom constraint. Assume that $A$ and $B$ infinitely update concurrently and this exact interleaving of messages continues. In this case $E$ is never able to update without causing glitches and is therefore stuck in a livelock. However, $E$ resolves this livelock as soon as $A$ or $B$ stop updating. In general, nodes in QPROP can livelock for graph topologies where two or more source nodes (e.g. $A$ and $B$) all propagate values to a single node in the graph (e.g. $D$) via two or more overlapping paths.

QPROP's livelocks resemble the "*duelling proposers*" [22] scenario for Paxos (i.e. two nodes alternately increase proposal numbers). Future work will focus on assessing whether randomisation [22, 24] (e.g. letting nodes await a random sleep timeout before handling a *change* message) could alleviate the livelock issue in practice.

## 8 Related Work

In Elm [11] each node in the dependency graph runs under its own thread of control. Moreover, each node has a number of queues which hold values propagated by predecessors. However, Elm relies on a *global event dispatcher* to provide new values to source nodes, which precludes elasticity and resilience.

A number of non-distributed reactive programming languages have recently been researched (e.g. FrTime [9] and Flapjax [21]). As discussed in Section 2 these languages topologically sort the dependency graph underlying reactive applications to guarantee glitch freedom. A similar approach is taken by synchronous reactive programming languages, such as Esterel [4], which rely on a scheduler to determine the order in which values propagate through the dependency graph.

*Globally asynchronous locally synchronous* (GALS) [3] systems provide a distributed approach to synchronous reactive programming. GALS discriminate between two kinds of systems: *exochronous* and *endo/isochronous* [17]. Endo/isochronous systems are defined by the fact that the system only relies on the values of signals, never on the presence/absence of the value of a signal. Concretely, this constraints the system to having a number of input signals for which one can infer at what pace they produce values. Using this classification, QPROP explicitly targets exochronous systems where multiple source signals might produce values at different and varying rates. To our knowledge current GALS systems solely target endo/isochronous systems [13, 25]. Although exochronous systems can be *endochronised*, this entails the addition of a centralised monitor or master clock to obtain the presence/absence of a signal value [3].

Quality-aware reactive programming (QUARP) [26] abstracts away the notion of glitches to a more general notion of *propagation quality*. This allows QUARP to implement decentralised glitch freedom as well as other propagation criteria (e.g. geographical location of nodes). QUARP and QPROP fundamentally differ in the glitch freedom guarantees they provide. In a nutshell, QUARP nodes only maintain the last propagated value for each of their predecessors. This value is overwritten each time the node receives a value from its associated predecessor. Nodes in QUARP can therefore livelock for *any* graph topology able to cause glitches (see Figure 3(A)). In contrast, QPROP nodes can never livelock for the topology shown in Figure 3(A). Moreover, QUARP does not guard against the issues which arise from dynamic topology changes (see Section 5). We refer the reader to Appendix C for a more detailed discussion regarding livelocks in QUARP and QPROP.

The work presented in [30] extends the synchronous reactive programming language Céu [29] with support for GALS systems. However, as stated in [30] it does not guarantee glitch freedom.

SID-UP [12] is a propagation algorithm specifically designed towards the development of decentralised distributed systems. However, as discussed in Section 2, SID-UP requires a central coordinator to support systems where source nodes update concurrently. This feature of SID-UP makes it unfit to deal with the reactive systems we envision.

Spreadsheet applications such as Microsoft's Excel essentially allow one to write asynchronous reactive code. Each cell in the spreadsheet can be seen as a signal which can depend on the values contained in other cells. Moreover, given Excel's multi-threaded capabilities [7] these updates can happen concurrently. A generalisation of this model to stream processing was introduced by [32]. However, glitches are avoided in this model by analysing the dependencies between cells. Such an analysis cannot be employed in a distributed context without resorting to a centralised entity. Moreover, this analysis would need to be re-run upon dynamic changes to the dependency graph incurring a substantial overhead.

AmbientTalk/R [8] is a reactive extension to the AmbientTalk [10] language. However, AmbientTalk/R only guards against local glitches (i.e. only signals residing on the same physical device are updated glitch freely).

## 9    Conclusion

Reactivity describes both certain kinds of programming languages as well as certain kinds of systems. Reactive programming languages provide constructs to declaratively implement event-driven applications. Reactive systems always respond to inputs in a timely fashion. For single-threaded non-distributed programs one can implement a reactive system using a reactive programming language. However, we observe that when moving to a distributed setting this is no longer the case. Using existing approaches, programmers are unable to implement reactive distributed systems using distributed reactive programming languages.

This lack of reactivity on behalf of current distributed reactive programming approaches stems from their propagation algorithms. We identify two key issues with these algorithms. First, they rely on central coordination to guarantee correctness (i.e. glitch freedom) of the distributed system. Second, they assume a lack of partial failures in the systems they support. In terms of reactive distributed systems [5, 6, 16] this entails that current DRP propagation algorithms are neither elastic nor resilient.

In this paper we propose two novel propagation algorithms, QPROP and its dynamic extension QPROP$^d$, which guarantee glitch-free propagation of values without resorting to centralised coordination. This decentralised design enables QPROP and QPROP$^d$ to support elastic systems. Moreover, QPROP and QPROP$^d$ embrace partial failures within distributed reactive applications. Nodes are able to crash without affecting the system as a whole. In other words, QPROP and QPROP$^d$ are resilient. We prove that QPROP and QPROP$^d$ are able to guarantee glitch freedom and show, through comparative benchmarks, that their decentralised design positively impacts reactivity. In conclusion QPROP and QPROP$^d$ bridge the gap between distributed reactive programming and reactive distributed systems. Programmers using frameworks or languages built atop QPROP or QPROP$^d$ are able to implement their reactive distributed systems using distributed reactive programming.

## A  Overview of Node Representation

■ **Table 1**  Overview of information held by each node in the dependency graph

| Abbreviation | Explanation |
|---|---|
| *DP* | Set containing references to the node's direct predecessors. |
| *DS* | Set containing references to the node's direct successors. |
| *I* | Dictionary containing the input set for each direct predecessor of the node. |
| *S* | Dictionary where the keys are references to source nodes and the values are sets of references to direct predecessor which are included in paths from the key source node to the node. |
| *U* | The node's update function. |
| *initVal* | The node's initial value. |
| *lastProp* | The node's last propagated value. |
| *self* | A reference to the node. |
| *clock* | The node's logical clock. |
| $B_r$ | Dictionary containing sets of propagation values for brittle predecessors (used by QPROP[d]). |

## B  Example Propagation

This section serves as an example of QPROP's workings. Consider an application comprised of five microservices as shown in Figure 10. The update functions of these microservices go as follows:

C  additions the values it receives from *A* and *B* (i.e. $C = A + B$).

D  subtracts the values it receives from *A* and *B* (i.e. $D = A - B$).

E  additions the values it receives from *C* and *D* (i.e. $E = C + D = (A + B) + (A - B) = 2 * A$).

Figure 10 provides an overview of the application's state as values propagate through the microservices. We discuss QPROP's behaviour at each time step.

t=0  The barrier phase completed. We assume that *A*, *B*, *C*, *D* and *E* respectively have 5, 3, 8, 2 and 10 as initial values. Each node stores its predecessors' initial propagation values in its *I* set. For example, both *C* and *D* store *A*'s initial value (i.e. $(A, 5, \{[A, 0]\}, 0)$) in their *I.A* sets (see Section 4.2.2 for an overview of the data contained in propagation values). Moreover, the figure also shows each node's *S* set. For example, *E*'s *S* set contains two entries: one which specifies that *C* and *D* propagate values originating from *A* and one which specifies that *C* and *D* propagate values originating from *B*. In other words, *E* can only update using values from *C* and *D* if the following holds:

$$U(val_C, val_D) \iff val_C.sClocks.A == val_D.sClocks.A \land val_C.sClocks.B == val_D.sClocks.B$$

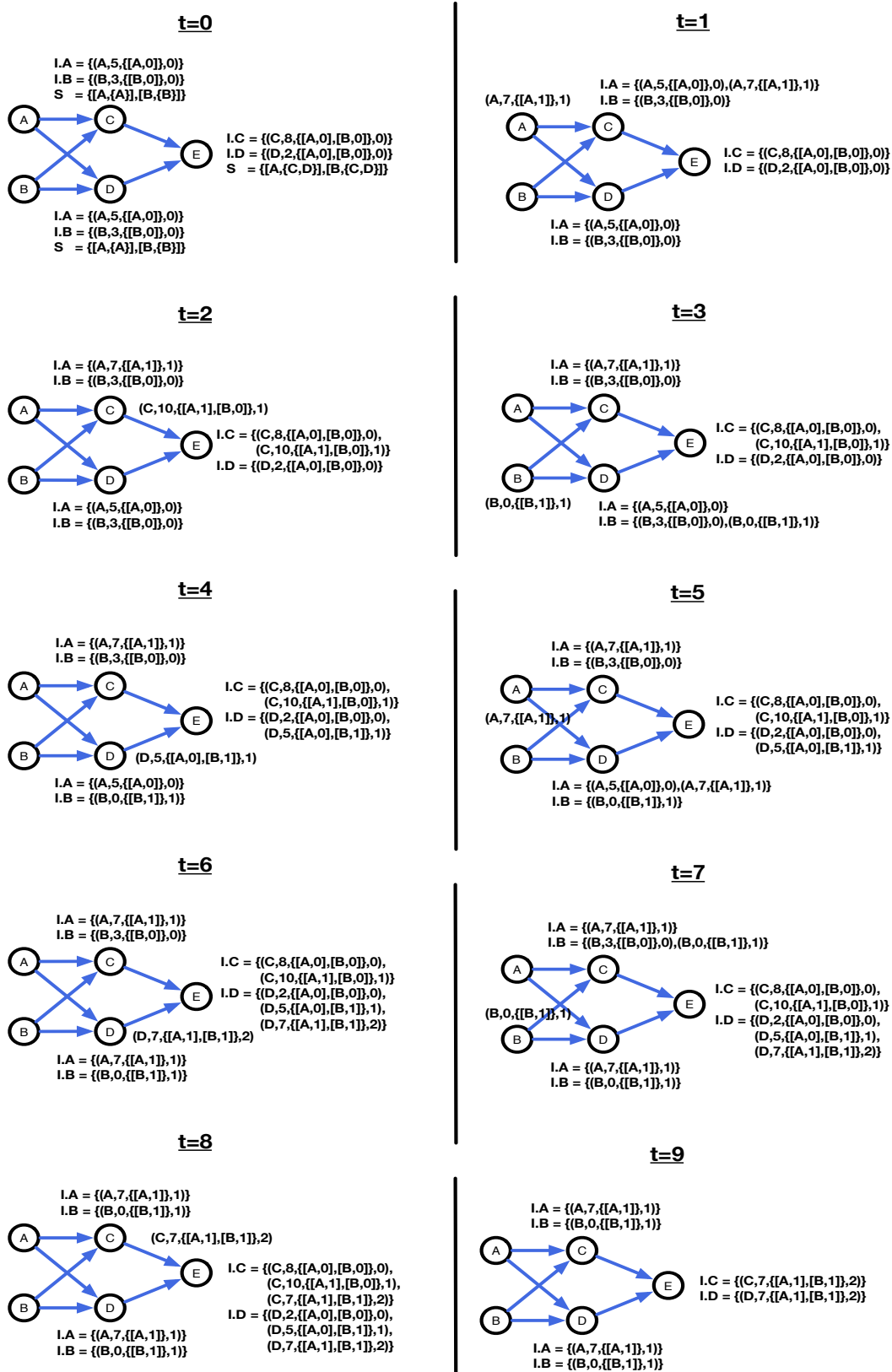

**Figure 10** Propagation of change with QPROP

**t=1** $A$ updates to 7 and propagates a new value $(A, 7, \{[A, 1]\}, 1)$ to its direct successors. At this point in time, only $C$ receives this new value and stores it in its $I.A$ set.

**t=2** Given that $C$ just received a new value it calculates the cross product between $\{(A, 7, \{[A, 1]\}, 1)\}$ and $I.B$. This results in a single set of arguments namely:

- $\{(A, 7, \{[A, 1]\}, 1), (B, 3, \{[B, 0]\}, 0)\}$

This set of arguments is trivially glitch free given that $\nexists [s, preds] \in S : A \in preds \wedge B \in preds$ (i.e. the values received from $A$ and $B$ do not originate from a common source). $C$ invokes its update lambda with 7 and 3 as arguments and propagates the resulting value (i.e. $(C, 10, \{[A, 1], [B, 0]\}, 1)$) to $E$ which stores it in its $I.C$ set. $C$ removes all values from its $I.A$ set which have an $fClock$ value smaller than 1 and all values from its $I.B$ set which have an $fClock$ value smaller than 0.

**t=3** $E$ calculates the cross product between $\{(C, 10, \{[A, 1], [B, 0]\}, 1)\}$ and $I.D$. This results in a single set of possible arguments for $E$:
$\{(C, 10, \{[A, 1], [B, 0]\}, 1), (D, 2, \{[A, 0], [B, 0]\}, 0)\}$. However, this set is not glitch free given that both arguments do not have equal clock values for $A$. $E$ therefore refrains from invoking its update function.

Meanwhile, $B$ updates to 0 and propagates this new value $(B, 0, \{[B, 1]\}, 1)$ to its direct successors. At this point in time, only $D$ receives this new value and stores it in its $I.B$ set.

**t=4** $D$ calculates the cross product between $\{(B, 0, \{[B, 1]\}, 1)\}$ and $I.A$. This results in a single set of glitch-free arguments namely:

- $\{(A, 5, \{[A, 0]\}, 0), (B, 0, \{[B, 1]\}, 1)\}$

$D$ invokes its update function with 5 and 0 as arguments and propagates the resulting value (i.e. $(D, 5, \{[A, 0], [B, 1]\}, 1)$) to $E$ which stores it in its $I.D$ set. $D$ removes all values from $I.A$ which have an $fClock$ value smaller than 0 and all values from $I.B$ which have an $fClock$ value smaller than 1.

**t=5** $E$ calculates the cross product between $\{(D, 5, \{[A, 0], [B, 1]\}, 1)\}$ and $I.C$. This results in two sets of possible arguments for $E$:

- $\{(C, 8, \{[A, 0], [B, 0]\}, 0), (D, 5, \{[A, 0], [B, 1]\}, 1)\}$
- $\{(C, 10, \{[A, 1], [B, 0]\}, 1), (D, 5, \{[A, 0], [B, 1]\}, 1)\}$

This first set is not glitch free given that both arguments do not have equal clock values for $B$. The second set is not glitch free either given that both arguments do not have equal clock values for $A$ nor $B$. $E$ therefore refrains from invoking its update function.

Meanwhile, $D$ receives the value propagated by $A$ at time $t=1$ and stores it in its $I.A$ set.

**t=6** $D$ calculates the cross product between $\{(A, 7, \{[A, 1]\}, 1)\}$ and $I.B$. This results in a single set of possible arguments: $\{(A, 7, \{[A, 1]\}, 1), (B, 0, \{[B, 1]\}, 1)\}$. $D$ invokes its update function with 7 and 0 as arguments and propagates the resulting value (i.e. $(D, 7, \{[A, 1], [B, 1]\}, 2)$ ) to $E$ which stores it in its $I.D$ set. $D$ removes all values from its $I.A$ and $I.B$ sets which have an $fClock$ value smaller than 1.

**t=7** $E$ calculates the cross product between $\{(D, 7, \{[A, 1], [B, 1]\}, 2)\}$ and $I.C$. This results in two sets of possible arguments for $E$:

- $\{(C,8,\{[A,0],[B,0]\},0),(D,7,\{[A,1],[B,1]\},2)\}$
- $\{(C,10,\{[A,1],[B,0]\},1),(D,7,\{[A,1],[B,1]\},2)\}$

This first set is not glitch free given that both arguments do not have equal clock values for $A$ nor $B$. The second set is not glitch free either given that both arguments do not have equal clock values for $B$. $E$ therefore refrains from invoking its update function.

Meanwhile, $C$ receives the value propagated by $B$ at time $t=3$ and stores it in its $I.B$ set.

**t=8** $C$ calculates the cross product between $\{(B,0,\{[B,1]\},1)\}$ and $I.A$. This results in a single set of glitch-free arguments namely: $\{(A,7,\{[A,1]\},1),(B,0,\{[B,1]\},1)\}$. $C$ invokes its update function with 7 and 0 as arguments and propagates the resulting value (i.e. $(C,7,\{[A,1],[B,1]\},2)$) to $E$ which stores it in its $I.C$ set. $C$ removes all values from its $I.A$ and $I.B$ sets which have an *fClock* value smaller than 1.

**t=9** $E$ calculates the cross product between $\{(C,7,\{[A,1],[B,1]\},2)\}$ and $I.D$. This results in three possible sets of arguments:

- $\{(C,7,\{[A,1],[B,1]\},2),(D,2,\{[A,0],[B,0]\},0)\}$
- $\{(C,7,\{[A,1],[B,1]\},2),(D,5,\{[A,0],[B,1]\},1)\}$
- $\{(C,7,\{[A,1],[B,1]\},2),(D,7,\{[A,1],[B,1]\},2)\}$

The first set is not glitch free given that both arguments do not have equal clock values for $A$ nor $B$. The second set is not glitch free either given that both arguments to do not have equal clock values for $A$. However, the last set of arguments fulfils $E$'s glitch freedom constraint. $E$ therefore invokes its update function with 7 and 7 resulting in 14. Note that $E$ therefore updates with twice the value of $A$'s update at time $t=1$, as prescribed by our example. $E$ removes all values from its $I.C$ and $I.D$ sets which have *fClock* values smaller than 2.

## C    Livelocks in QPROP and QUARP

This section provides two concrete examples of livelocks in distributed reactive programming. In the first example all nodes run the QPROP algorithm while nodes in the second example run the QUARP [26] algorithm.

### C.1    Livelocks in QPROP

Figure 11 provides an overview of the application's state as values propagate through the microservices. We discuss QPROP's behaviour and how $E$ livelocks at each time step.

**t=0** The barrier phase completed. Each node stores its predecessors' initial propagation values in its $I$ set. We remind the reader that $E$ can only update using values from $C$ and $D$ if the following holds:

$$U(val_C, val_D) \iff val_C.sClocks.A == val_D.sClocks.A \land val_C.sClocks.B == val_D.sClocks.B$$

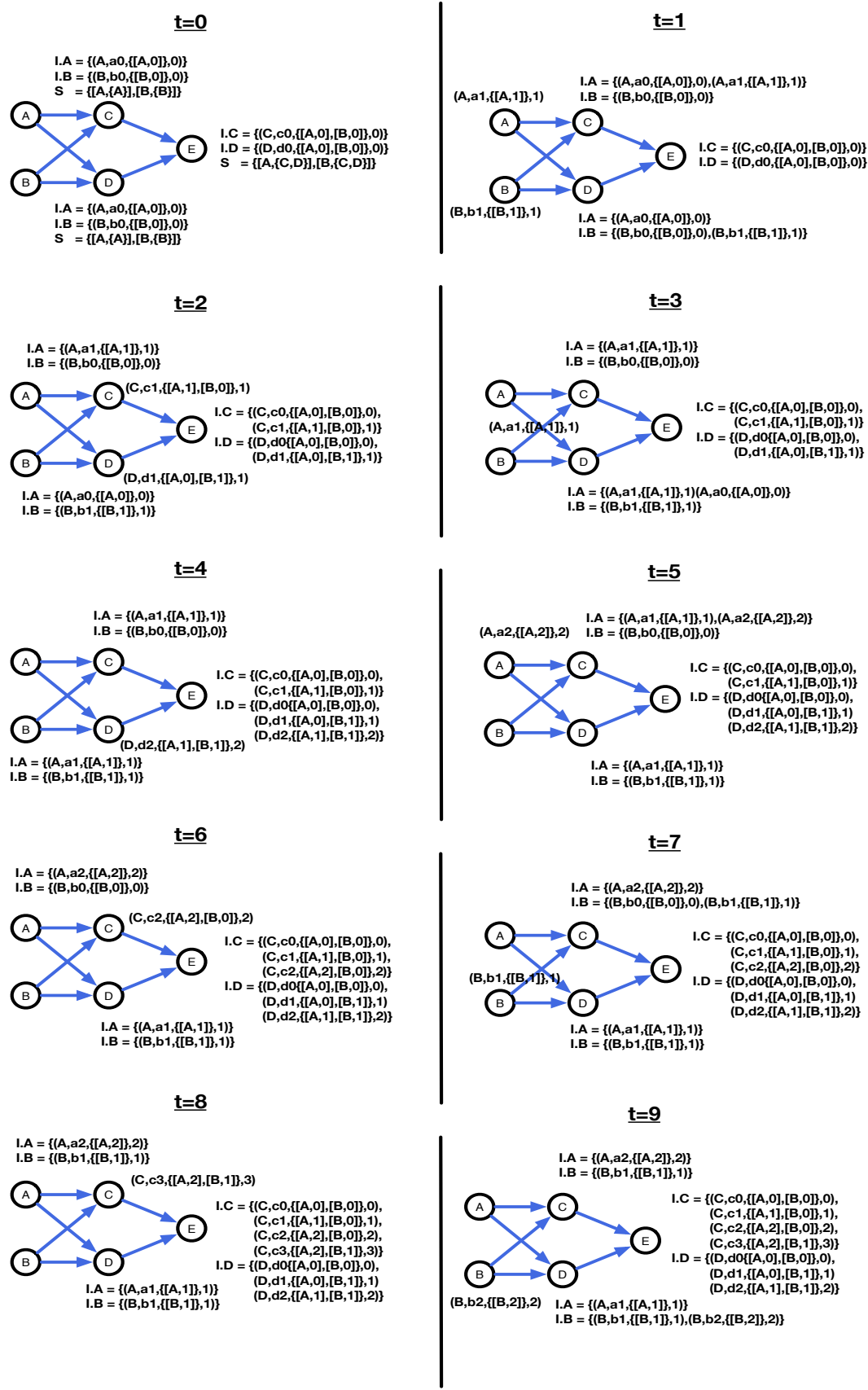

**Figure 11** Livelocks in QPROP

**t=1** *A* updates to *a1* and propagates a new value $(A, a1, \{[A, 1]\}, 1)$ to its direct successors. At this point in time, only *C* receives this new value and stores it in its *I.A* set. Moreover, *B* updates to *b1* and propagates a new value $(B, b1, \{[B, 1]\}, 1)$ to its direct successors. At this point in time, only *D* receives this new value and stores it in its *I.B* set.

**t=2** *C* calculates the cross product between $\{(A, a1, \{[A, 1]\}, 1)\}$ and *I.B*. This results in a single set of possible arguments: $\{(A, a1, \{[A, 1]\}, 1), (B, b0, \{[B, 0]\}, 0)\}$. *C* invokes its update function with these values as arguments and propagates the resulting value (i.e. $(C, c1, \{[A, 1], [B, 0]\}, 1)$ ) to *E* which stores it in its *I.C* set. *C* removes all values from its *I.A* set which have an *fClock* value smaller than 1 and all values from its *I.B* set which have an *fClock* value smaller than 0.

*D* calculates the cross product between $\{(B, b1, \{[B, 1]\}, 1)\}$ and *I.A*. This results in a single set of possible arguments: $\{(A, a0, \{[A, 0]\}, 0), (B, b1, \{[B, 1]\}, 1)\}$. *D* invokes its update function with these values as arguments and propagates the resulting value (i.e. $(D, d1, \{[A, 0], [B, 1]\}, 1)$ ) to *E* which stores it in its *I.D* set. *D* removes all values from its *I.A* set which have an *fClock* value smaller than 0 and all values from its *I.B* sets which have an *fClock* value smaller than 1.

**t=3** Upon receiving *C*'s new value (i.e. before receiving *D*'s new value) *E* calculates the cross product between $\{(C, c1, \{[A, 1], [B, 0]\}, 1)\}$ and *I.D*. This results in a single sets of possible arguments for *E*:

- $\{(C, c1, \{[A, 1], [B, 0]\}, 1), (D, d0, \{[A, 0], [B, 0]\}, 0)\}$

This set is not glitch free given that both arguments do not have equal clock values for *A*. *E* therefore refrains from invoking its update function. Upon receiving *D*'s new value (i.e. after receiving *C*'s new value) *E* calculates the cross product between $\{(D, d1, \{[A, 0], [B, 1]\}, 1)\}$ and *I.C* This results in two sets of possible arguments for *E*:

- $\{(C, c0, \{[A, 0], [B, 0]\}, 1), (D, d1, \{[A, 0], [B, 1]\}, 1)\}$
- $\{(C, c1, \{[A, 1], [B, 0]\}, 1), (D, d1, \{[A, 0], [B, 1]\}, 1)\}$

None of these sets are glitch free, *E* refrains from invoking its update function. Meanwhile, *D* receives the value propagated by *A* at time *t=1* and stores it in its *I.A* set.

**t=4** *D* calculates the cross product between $\{(A, a1, \{[A, 1]\}, 1)\}$ and *I.B*. This results in a single set of possible arguments:

- $\{(A, a1, \{[A, 1]\}, 1), (B, b1, \{[B, 1]\}, 1)\}$

*D* invokes its update function with these values as arguments and propagates the resulting value (i.e. $(D, d2, \{[A, 1], [B, 1]\}, 2)$ ) to *E* which stores it in its *I.D* set. *D* removes all values from its *I.A* and *I.B* set which have an *fClock* value smaller than 1.

**t=5** *E* calculates the cross product between $\{(D, d2, \{[A, 1], [B, 1]\}, 2)\}$ and *I.C*. This results in two sets of possible arguments for *E*:

- $\{(C, c0, \{[A, 0], [B, 0]\}, 1), (D, d2, \{[A, 1], [B, 1]\}, 2)\}$
- $\{(C, c1, \{[A, 1], [B, 0]\}, 1), (D, d2, \{[A, 1], [B, 1]\}, 2)\}$

None of these sets are glitch free, $E$ refrains from invoking its update function. Meanwhile, $A$ updates to $a2$ and propagates a new value $(A, a2, \{[A, 2]\}, 2)$ to its direct successors. At this point in time, only $C$ receives this new value and stores it in its $I.A$ set.

**t=6** $C$ calculates the cross product between $\{(A, a2, \{[A, 2]\}, 2)\}$ and $I.B$. This results in a single set of possible arguments:

- $\{(A, a2, \{[A, 2]\}, 2), (B, b0, \{[B, 0]\}, 0)\}$

$C$ invokes its update function with these values as arguments and propagates the resulting value (i.e. $(C, c2, \{[A, 2], [B, 0]\}, 2)$ ) to $E$ which stores it in its $I.C$ set. $C$ removes all values from its $I.A$ set which have an *fClock* value smaller than 2 and all values from its $I.B$ set which have an *fClock* value smaller than 0.

**t=7** $E$ calculates the cross product between $\{(C, c2, \{[A, 2], [B, 0]\}, 2)\}$ and $I.D$ This results in three sets of possible arguments for $E$:

- $\{(C, c2, \{[A, 2], [B, 0]\}, 2), (D, d0, \{[A, 0], [B, 0]\}, 0)\}$
- $\{(C, c2, \{[A, 2], [B, 0]\}, 2), (D, d1, \{[A, 0], [B, 1]\}, 1)\}$
- $\{(C, c2, \{[A, 2], [B, 0]\}, 2), (D, d2, \{[A, 1], [B, 1]\}, 2)\}$

None of these sets are glitch free, $E$ refrains from invoking its update function. Meanwhile, $C$ receives the value propagated by $B$ at time $t=1$ and stores it in its $I.B$ set.

**t=8** $C$ calculates the cross product between $\{(B, b1, \{[B, 1]\}, 1)\}$ and $I.C$. This results in a single set of possible arguments:

- $\{(A, a2, \{[A, 2]\}, 2), (B, b1, \{[B, 1]\}, 1)\}$

$C$ invokes its update function with these values as arguments and propagates the resulting value (i.e. $(C, c3, \{[A, 2], [B, 1]\}, 3)$ ) to $E$ which stores it in its $I.C$ set. $C$ removes all values from its $I.A$ set which have an *fClock* value smaller than 2 and all values from its $I.B$ set which have an *fClock* value smaller than 1.

**t=9** $E$ calculates the cross product between $\{(C, c3, \{[A, 2], [B, 1]\}, 3)\}$ and $I.D$ This results in three sets of possible arguments for $E$:

- $\{(C, c3, \{[A, 2], [B, 1]\}, 3), (D, d0, \{[A, 0], [B, 0]\}, 0)\}$
- $\{(C, c3, \{[A, 2], [B, 1]\}, 3), (D, d1, \{[A, 0], [B, 1]\}, 1)\}$
- $\{(C, c3, \{[A, 2], [B, 1]\}, 3), (D, d2, \{[A, 1], [B, 1]\}, 2)\}$

None of these sets are glitch free, $E$ refrains from invoking its update function. $B$ updates to $b2$ and propagates a new value $(B, b2, \{[B, 2]\}, 2)$ to its direct successors. At this point in time, only $D$ receives this new value and stores it in its $I.B$ set.

The interleaving of messages for this example ensure that $C$ and $D$ never update using the same values for $A$ and $B$. As a result, $E$ is never able to find a set of arguments which do not cause glitches. Repeating this interleaving of messages therefore ensures that $E$ remains in a livelock.

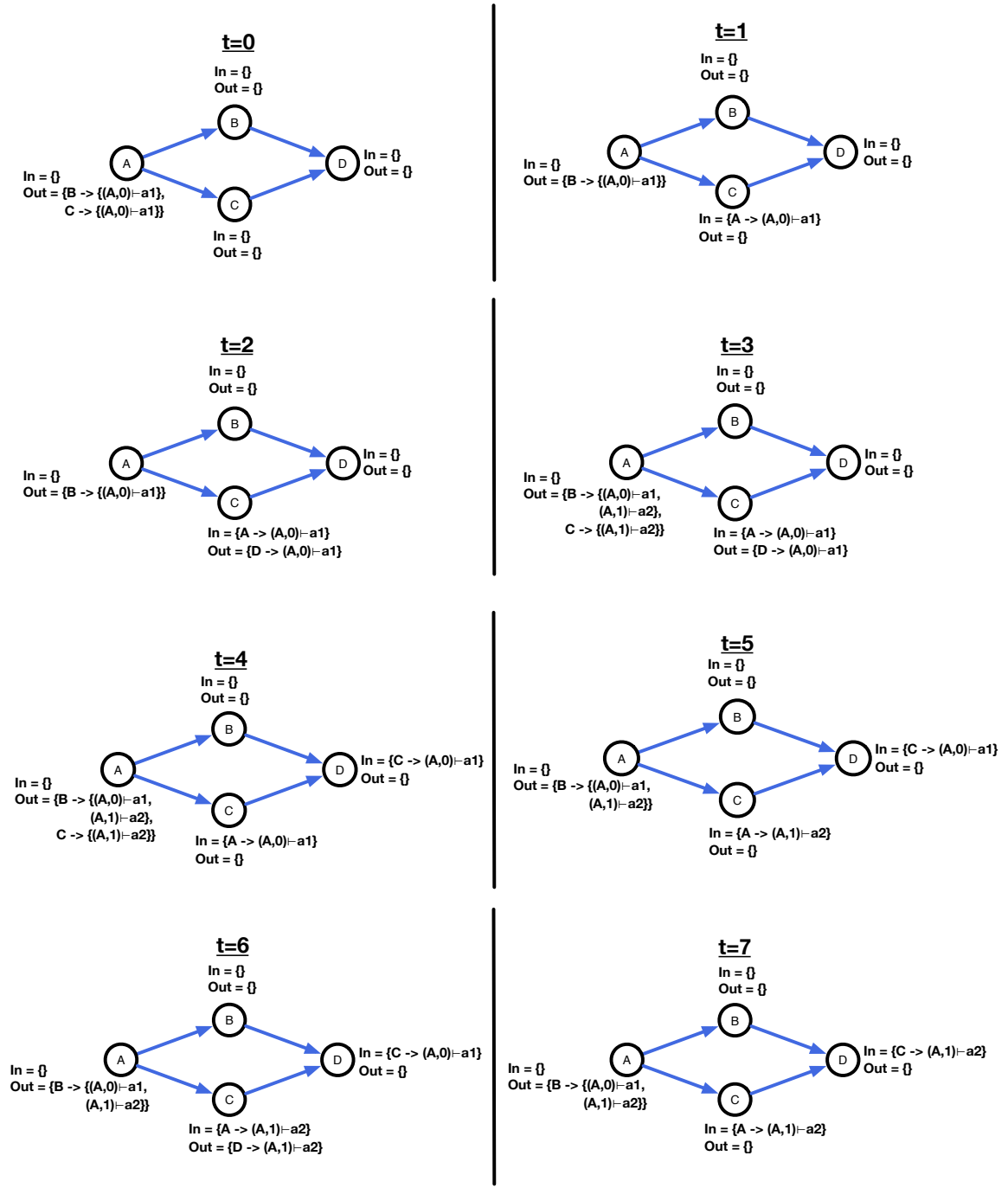

**Figure 12** Livelocks in QUARP

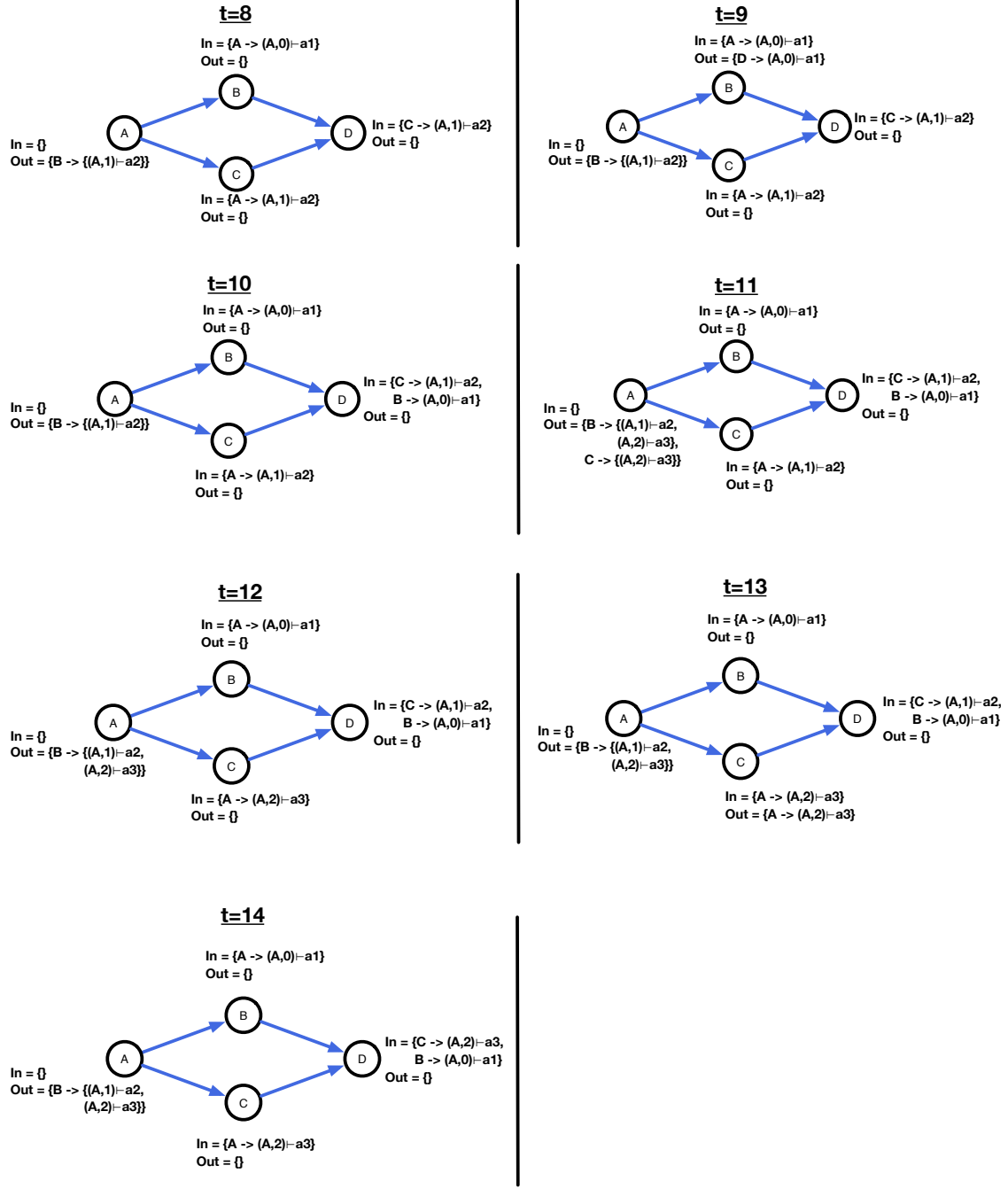

**Figure 13** Livelocks in QUARP

## C.2 Livelocks in QUARP

Figure 12 and Figure 13 provide an overview of the application's state as values propagate through the microservices. Based on QUARP's operational semantics [26] we discuss how updates are propagated through the distributed dependency graph.

**t=0** Source node $A$ adds $a1$ with counter 0 to $out_A$ (src rule).

**t=1** $A$ transfers $a1$ to $in_C$ (rcv rule).

**t=2** $C$ evaluates and adds the result to $out_C$ (pub rule).

**t=3** Source node $A$ adds $a2$ with counter 1 to $out_A$ (src rule).

**t=4** $C$ transfers $a1$ to $in_D$ (rcv rule). $D$ is unable to evaluate given that it is not active (i.e. all of its input buffers do not contain a value).

**t=5** $A$ transfers $a2$ to $in_C$ (rcv rule).

**t=6** $C$ evaluates and adds the result to $out_C$ (pub rule).

**t=7** $C$ transfers $a2$ to $in_D$ (rcv rule). $D$ is unable to evaluate given that it is not active.

**t=8** $A$ transfers $a1$ to $in_B$ (rcv rule).

**t=9** $B$ evaluates and adds the result to $out_B$ (pub rule).

**t=10** $B$ transfers $a1$ to $in_D$ (rcv rule). $D$ is unable to evaluate given that the minimal glitch freedom quality is not met (i.e. $A = A \rightarrow 1 = 0$ is false).

**t=11** $A$ adds $a3$ with counter 2 to $out_A$ (src rule)).

**t=12** $A$ transfers $a3$ to $in_C$ (rcv rule).

**t=13** $C$ evaluates and adds the result to $out_C$ (pub rule).

**t=14** $C$ transfers $a3$ to $in_D$ (rcv rule). $D$ is unable to evaluate given that the minimal glitch freedom quality is not met (i.e. $A = A \rightarrow 2 = 0$ is false).

In this example $C$ propagates values at a higher pace than $B$ (e.g. the link between $A$ and $B$ suffers from higher latency than the link between $A$ and $C$). Nodes in QUARP systematically overwrite the previous value received from a predecessor. In other words, each time $E$ receives a new value from $C$ it overwrites the previous value received from $C$. $E$ will therefore never be able to find a pair of values for which it can update without causing glitches. As such, $E$ is stuck in a livelock.

## C.3 Comparing QUARP and QPROP

The fundamental difference between QUARP and QPROP is characterised by the kind of topologies for which they might livelock.

QUARP nodes are vulnerable to livelocks for all graphs which can possibly exhibit a glitch. The topology of such graphs (see Figure 14(A)) is characterised by a single source node (i.e. $A$) which propagates values to a single node in the graph (i.e. $C$) via more than one path. All such topologies can cause QUARP to let $C$ livelock (see Section C.2 for an example). In contrast, QPROP nodes cannot livelock for such topologies given that QPROP nodes do not overwrite values received from predecessors.

The topology of graphs for which QPROP nodes can livelock (see Figure 14(B)) is characterised by two or more source nodes (i.e. $A$ and $B$) which all propagate values to a single node in the graph (i.e. $D$) via two or more overlapping paths (see Section C.1 for an example).

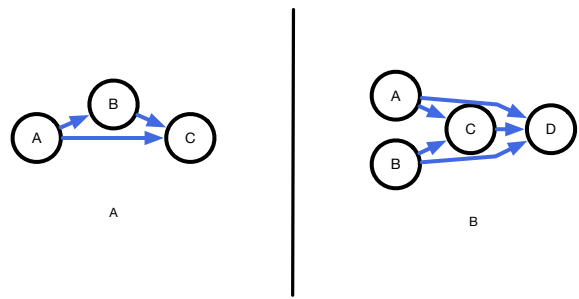

**Figure 14** A) minimal dependency graph topology for which QUARP livelocks. B) minimal dependency graph topology for which QUARP and QPROP livelock.

It is important to note that this second kind of topology is a specialisation of the first: the subgraphs formed by the nodes $\{A, C, D\}$ and $\{B, C, D\}$ are both examples of the first kind of topology. As such, all graphs for which QPROP nodes might livelock contain at least two sub-graphs for which QUARP nodes might livelock. In other words, QUARP nodes are vulnerable to livelocks for all topologies for which QPROP nodes are vulnerable to livelocks but not vice versa. However, a QPROP node will eventually run out of memory in case of a livelock as it stores all values received from predecessors. This is not the case for QUARP nodes, given that these overwrite previously received values.

## D Dynamic Graph Topology Changes

The most crucial parts of QPROP and QPROP$^d$ are presented in Section 4.2 and Section 5 respectively. In this section we detail the intentional graph topology changes of QPROP$^d$ which are omitted from Section 5, namely nodes dynamically leaving or joining the dependency graph and removing dependencies between nodes.

### D.1 Addendum to Algorithm 4

Algorithm 5 contains the predicates and the *MoveToI* function omitted from Algorithm 4 for the sake of brevity. It is important to note that the predicates are to be considered as macros (i.e. they expand when the node runs Algorithm 4). In other words, the

---

**ALGORITHM 5:** Pre-propagation Predicates and MoveToI

1  **Predicate** $isBrittle(pred) = [pred, \_] \in B_r$
2  **Predicate** $hasBrittleSibling(pred) = \exists[s, dps] \in S, \exists pred_{brittle} \in dps : pred \in dps \wedge isBrittle(pred_{brittle})$
3  **Predicate** $isBrittleSibling(pred_{brittle}, pred) = isBrittle(pred_{brittle}) \wedge \exists[s, dps] \in S : pred_{brittle} \in dps \wedge pred \in dps$
4  **Predicate** $synchronised(pred_{brittle}) = \forall pred \in DP, \forall[s, dps] \in S : isBrittle(pred_{brittle}) \wedge pred \in dps \wedge pred_{brittle} \in dps \implies$
5      $B_r.pred_{brittle}.first().sClocks.s - I.pred.first().sClocks.s \leq 1$
6  **Function** $MoveToI(pred_{brittle})$:
7      **if** $[pred_{brittle}, \_] \in I$ **then**
8          $\mid \quad I.pred_{brittle} = I.pred_{brittle} \cup B_r.pred_{brittle}$
9      **else**
10         $\mid \quad I = I \cup \{[pred_{brittle}, B_r.pred_{brittle}]\}$
11     **end**
12     $B_r = B_r \setminus \{[pred_{brittle}, \_]\}$
13     **return**
14

---

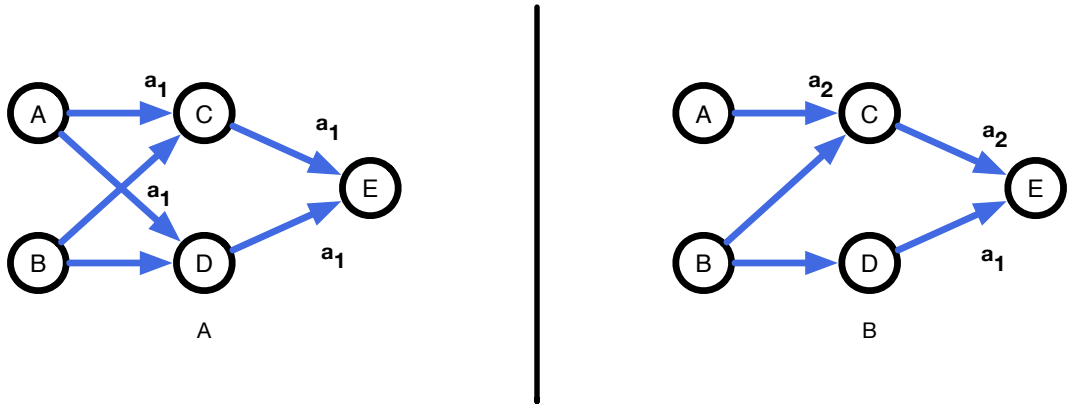

■ **Figure 15**   (A)Dependency graph before dynamic removal of a dependency between $A$ and $D$. Source $A$ propagates value $a_1$.(B)Dependency graph after dynamic removal of a dependency between $A$ and $D$. Source $A$ propagates value $a_2$.

free variables in the predicates (e.g. $B_r$ in *isBrittle*) are bound to the corresponding node elements upon expansion.

### D.2  Dynamic Dependency Removal

Figure 15(A) and (B) respectively show the state of a dependency graph before and after removing the dependency between $A$ and $D$. As is the case for the addition of a new dependency (see Section 5.1), QPROP[d]'s task is to guarantee glitch freedom during the removal operation while values are propagating through the dependency graph.

Assume $A$ propagates $a_1$ before the dependency between $A$ and $D$ is removed. Furthermore, assume $E$ is able to update itself using these two values (see Figure 15(A)). At this point $E$'s $S$ dictionary contains two entries: $[A, \{C, D\}]$ and $[B, \{C, D\}]$ In Figure 15(B) the dependency between $A$ and $D$ has been removed. This entails that the $A$ entry in $E$'s $S$ dictionary now looks as follows: $[A, \{C\}]$. Assume $A$ propagates $a_2$ to $C$ (given that $D$ is no longer a successor of $A$). According to the *change* Handler (see Section 4.2.4), $E$ now only needs to ensure that it uses arguments for which the $B$ clock times are equal. This would lead $E$ to update itself using $a_1$ from $D$ and $a_2$ from $C$, which constitutes a glitch.

This problem arises whenever a node must modify an existing entry in its $S$ dictionary. In our example, $E$ removes $D$ from $[A, \{C, D\}]$. In other words, all values stored by $E$ in $I.D$ have become stale given that $D$ no longer propagates values which originate from $A$. To avoid this issue, $E$ must therefore remove all values received from $D$ in $I.D$.

---

**ALGORITHM 6:** Dynamic dependency removal

---

**arguments :** A predecessor to remove $pred$
1  $sources = await\ pred \leftarrow remSucc(self)$
2  $I = I \setminus \{I.pred\}$
3  $DP = DP \setminus \{pred\}$
4  $await\ self \leftarrow remSources(pred, sources)$
5  **if** $|DP| == 0$ **then**
6      **foreach** $succ \in DS$ **do**
7          $await\ succ \leftarrow addSource(self, self)$
8      **end**
9  **end**

---

**Handler** remSucc(succ)

---

1  $DS = DS \setminus \{succ\}$
2  **if** $|DP| == 0$ **then**
3      **return** $\{self\}$
4  **else**
5      $allSources = \{s | [s, \_] \in S\}$
6      **return** $allSources$
7  **end**

---

**Handler** remSources(from,sources)

---

1  $removed = \{\}$
2  **foreach** $source \in sources$ **do**
3      $S.source = S.source \setminus \{from\}$
4      **if** $S.source == \emptyset$ **then**
5          $S = S \setminus \{[source, \{\}]\}$
6          $removed = removed \cup \{source\}$
7      **else**
8          $I.from = \{\}$
9      **end**
10 **end**
11 **foreach** $succ \in DS$ **do**
12     $await\ succ \leftarrow remSources(self, removed)$
13 **end**

---

**Handler** addSource(from,source)

---

1  **if** $[source, \_] \in S$ **then**
2      $S.source = S.source \cup \{from\}$
3  **else**
4      $S = S \cup \{[source, \{from\}]\}$
5  **end**
6  **foreach** $succ \in DS$ **do**
7      $await\ succ \leftarrow addSource(self, source)$
8  **end**

---

A node $n$ dynamically removing a dependency to a predecessor $pred$ runs Algorithm 6. In a first step, $n$ informs $pred$ that it is removing the dependency by sending the *remSucc* message. Furthermore, $n$ removes $pred$'s input set from $I$ and updates its direct predecessors $DP$. Upon receiving the *remSucc* message (see the remSucc Handler) $pred$ removes $n$ from its set of direct successors (i.e. $DS$) and returns a set with all sources which are able to reach it. $n$ uses this set as an argument to the *remSources* message which it sends to itself. The remSources Handler will recursively update the topological information held by $n$ and all its direct and downstream successors.

For our example depicted in Figure 15 $D$ will send itself the *remSources* message with $A$ as arguments for both *from* and *sources*. As a result, $D$ will remove $A$ from the $[A, \{A\}]$ entry in $S$ (see line 3). Given that $S.A$ is now empty, $D$ removes the entire entry from $S$ (see lines 5 and 6). In other words, $D$ will no longer propagate values originating from $A$ and notifies $E$ of this fact by recursively sending the *remSources* message to $E$ using $D$ and $A$ as arguments. Upon receiving the message, $E$ removes $D$ from $[A, \{C, D\}]$ in $S$. Given that $E$ can still receive values originating from $A$ through $C$ it must empty $D$'s input set (see line 8) in order to avoid the aforementioned glitch.

Once all of $n$'s direct and downstream successors have finished updating, $n$ needs to ensure that it didn't become a source node by removing the dependency with $pred$ (i.e. $pred$ was its only direct predecessor). If $n$ did become a source, it notifies its direct and

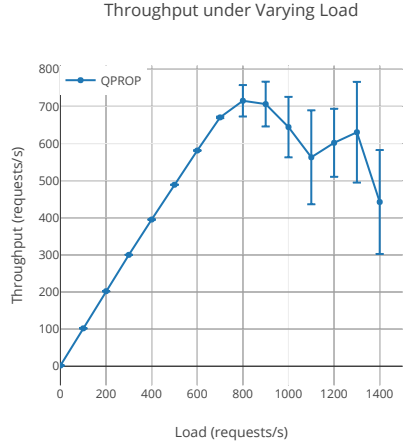

■ **Figure 16** QPROP's maximum throughput for the fleet management application. Error bars indicate the 95 % confidence interval downstream successors of this fact by sending the *addSource* message. In a nutshell, the addSource Handler updates a node's *S* set based on the newly reachable source.

### D.3 Dynamic Node Addition and Removal

| **ALGORITHM 7:** Dynamic node addition | **ALGORITHM 8:** Dynamic node removal |
|---|---|
| 1 **foreach** *pred* ∈ *DP* **do** | 1 **foreach** *pred* ∈ *DP* **do** |
| 2 │ /*Proceed with Algorithm 3*/ | 2 │ /*Proceed with Algorithm 6*/ |
| 3 **end** | 3 **end** |
| 4 **foreach** *succ* ∈ *DS* **do** | 4 **foreach** *succ* ∈ *DS* **do** |
| 5 │ /*Let succ Proceed with Algorithm 3*/ | 5 │ /*Let succ Proceed with Algorithm 6*/ |
| 6 **end** | 6 **end** |

Adding a node *n* dynamically to a dependency graph is equivalent to letting *n* sequentially run Algorithm 3 with each of its direct predecessors as arguments. Moreover, each of *n*'s direct successors runs Algorithm 3 with *n* as argument. Similarly, removing a node dynamically from a dependency graph follows the same pattern using Algorithm 6.

## E  Additional Benchmark Results

### E.1 Use Case Comparison

In Section 6.1 we compare QPROP's throughput with SID-UP's throughput for our fleet management application. The results show that SID-UP's maximum throughput is 100 requests per second, while QPROP is able to scale up to Emixis' daytime load of 300 requests per second. In order to measure QPROP's maximum throughput for our use case we vary the request load to 1400 requests per second. Figure 16 shows the

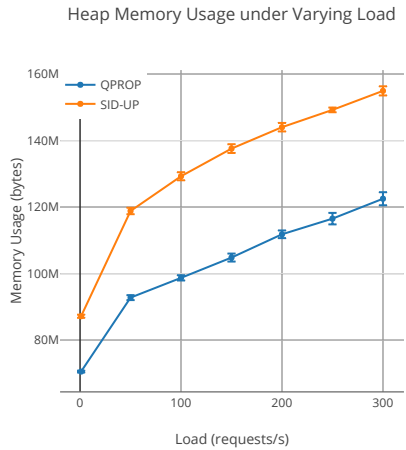

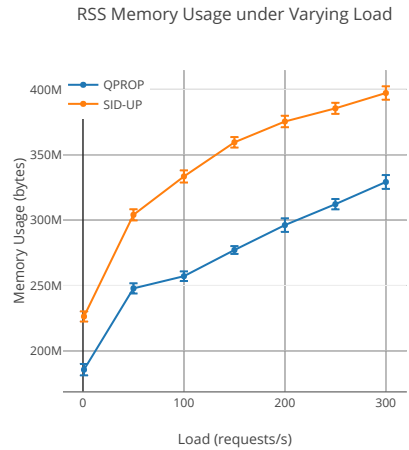

■ **Figure 17** Heap memory usage across services under varying loads. Error bars indicate the 95 % confidence interval

■ **Figure 18** Rss memory usage across services under varying loads. Error bars indicate the 95 % confidence interval

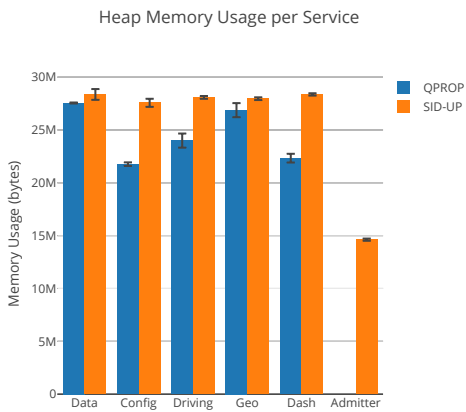

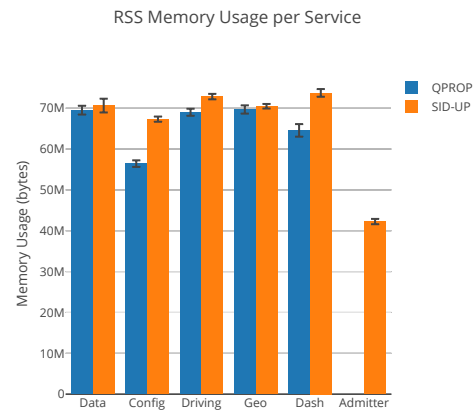

■ **Figure 19** Heap memory usage per service for a load of 300 requests per second. Error bars indicate the 95 % confidence interval

■ **Figure 20** RSS memory usage per service for a load of 300 requests per second. Error bars indicate the 95 % confidence interval results for this experiment. QPROP is roughly able to handle 700 requests per second, after which throughput becomes negatively affected by the increasing load.

Figure 17 and Figure 18 show the results for the memory measurements. SID-UP suffers from an overhead for both heap memory usage as well as RSS memory usage. To understand this overhead, consider Figure 19 and Figure 20 which detail the heap memory usage and RSS memory usage per service for a load of 300 requests per second. The services running QPROP consistently use less heap and RSS memory,

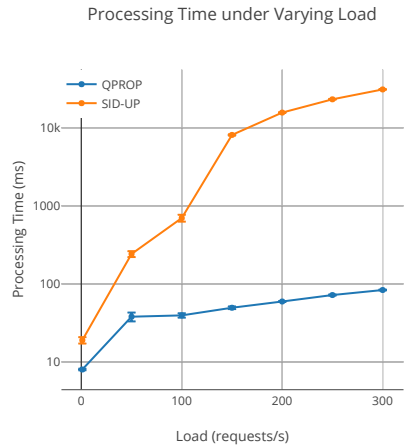

■ **Figure 21** Request processing time under varying load. Error bars indicate the 95 % confidence interval although this could be attributed to implementation differences. The major difference between both approaches comes from the fact that SID-UP requires an additional *admitter* service to coordinate updates to the application.

Figure 21 shows the request processing times for both approaches under varying load. Although QPROP suffers from a latency overhead, one clearly sees that SID-UP suffers from a much larger processing time overhead. The reason for this overhead is that a request can only be handled by SID-UP once the previous request has been handled. In contrast, QPROP allows our fleet management application to handle requests in parallel.

### E.2 General Comparison

This section contains the benchmark results for the larger microservice system which are omitted from Section 6.2.

Figure 22 shows the latency results comparing QPROP and SID-UP . Figure 23 shows the latency results comparing QPROP[d] and SID-UP under a load of 100 requests per second. As is the case for the fleet management application, QPROP introduces an overhead with regards to latency. Dynamic topology changes only seem to impact QPROP[d]'s latency periodically. More precisely, in QPROP[d] a topology change will render the part of the dependency graph affected by the change unresponsive until the change completes. As a result, our benchmarks show an increase in outliers while the average latency remains roughly similar across the benchmarks. In contrast, SID-UP's latency is unaffected by load or dynamic topology changes. The reason for this phenomenon is explained in Section 6.1. SID-UP's latency is unaffected by dynamic topology changes for essentially the same reason. A change is only performed on the dependency graph whenever the previous change has completed or the previous propagation value has traversed the graph.

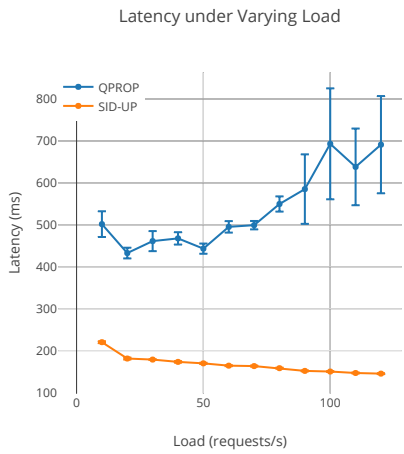

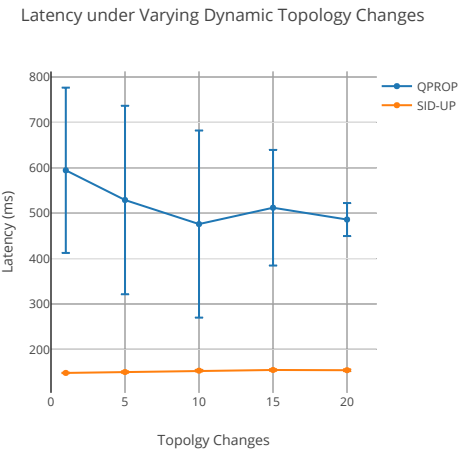

**Figure 22** Latency under varying loads. Error bars indicate the 95 % confidence interval

**Figure 23** Latency under varying dynamic topology changes. Error bars indicate the 95 % confidence interval

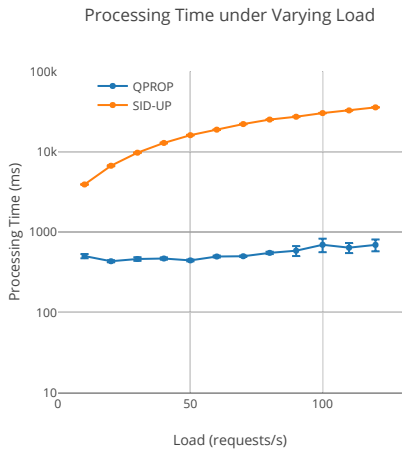

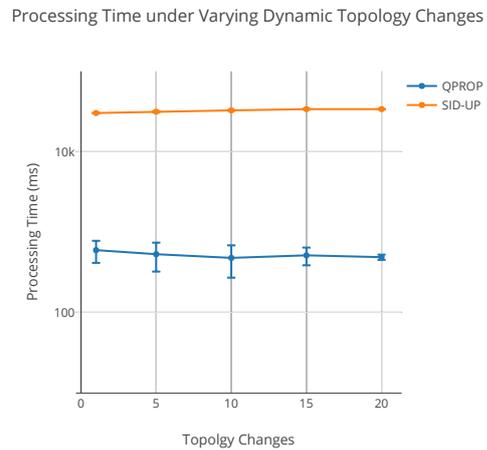

**Figure 24** Request processing time under varying load. Error bars indicate the 95 % confidence interval

**Figure 25** Request processing time under varying dynamic topology changes. Error bars indicate the 95 % confidence interval

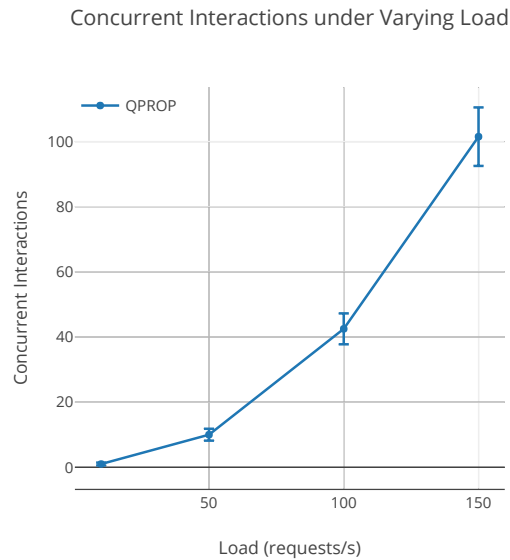

■ **Figure 26**  Concurrent interactions under varying loads. Error bars indicate the 95 % confidence interval

Figure 24 shows the average processing time per request. As is the case for the fleet management application, SID-UP suffers from a significant overhead as load increases. Figure 25 shows how processing times are affected by dynamic topology changes, both systems are put under a static load of 100 requests per second. The results show that topology changes hardly affect processing times.

### E.3  Concurrent Interactions

QPROP and QPROP[d] allow updates to source nodes to concurrently traverse the dependency graph. As a result it can happen that a node depending on two source nodes only updates once as a result of both source nodes updating. We call this phenomenon *concurrent interactions*.

We measure the amount of concurrent interactions for the larger, artificial, microservice system. Regular benchmarks are stopped whenever the sink nodes processed the given amount of load (see Section 6). For the *concurrent interactions* benchmarks we stop the benchmark as soon as the source nodes have produced the given amount of load. The difference between the generated amount of load by the source nodes and processed amount of load (i.e. updates) by the sink nodes allow us to measure the amount of concurrent interactions.

Figure 26 shows the amount of concurrent interactions as load increases. As the load increases the opportunities for multiple source nodes to update concurrently and for nodes in the dependency graph to receive partial updates increases. Hence, the amount of concurrent interactions increase as well.

## F  Glitch Freedom, Termination and Eventual Consistency

Informally, an application is glitched if it has only *partially* been updated as the result of a change in one of the event sources (i.e. one of the source nodes in the underlying dependency graph). Such a partial update is the result of an incorrect traversal of the dependency graph by the propagation algorithm. In other words, the algorithm has updated a certain node before it updated all of its predecessors in the dependency graph.

To aid us in proving [3] that QPROP is glitch free we introduce the following definitions:

**Definition 1 (Dependency Graph)**  *A dependency graph is a pair $(N, E)$ where $N$ is the set of nodes and $E$ is a set of pairs denoting directed edges between nodes in $N$. This graph consists of three types of nodes:*

- *Source nodes : $\{n_{so} \in N | \nexists n \in N : (n, n_{so}) \in E\}$.*
- *Intermediate nodes : $\{n_i \in N | \exists n_1, n_2 \in N : (n_1, n_i) \in E \wedge (n_i, n_2) \in E\}$.*
- *Sink nodes : $\{n_s \in N | \nexists n \in N : (n_s, n) \in E\}$*

**Definition 2 (Path)**  *$P(x, y)$ denotes the existence of a path in a dependency graph $(N, E)$ between nodes $x$ and $y$. In other words :*

$$P(x, y) \iff (x, y) \in E \vee \exists z \in N : P(x, z) \wedge P(z, y).$$

**Definition 3 (Propagation Path)**  *The set of reachable nodes starting from a source node $n_{so}$ is a partially ordered set: $PP_{n_{so}} = \{n_{so}\} \cup \{n \in N | P(n_{so}, n)\}$. The order of the nodes in this set is defined as follow: $\forall n_1, n_2 \in PP_{n_{so}} : n_1 > n_2 \iff P(n_1, n_2)$. We say that $PP_{n_{so}}$ is $n'_{so}s$ propagation path.*

**Definition 4 (Precedence)**  *We define that a node $x$ directly precedes a node $y$ in a dependency graph $(N, E)$ as follows: $x >> y \iff \exists(x, y) \in E$. Similarly, we define that a node $x$ directly succeeds a node $y$ in a dependency graph $(N, E)$ as follows: $x << y \iff \exists(y, x) \in E$ Note that both relations are non-transitive.*

### F.1  Exploration Correctness

QPROP's exploration phase guarantees that a node $n$'s $S_n$ dictionary contains entries which map each source node $n_{so}$ able to reach $n$ onto $n$'s direct predecessors $DP_n = \{dp \in N | dp >> n\}$. More precisely:

**Theorem 1 (Exploration Correctness)**

$$\forall n \in N, \forall dp_i \in DP_n : \exists P(n_{so}, dp_i) \iff \exists [n_{so}, \{..., dp_i, ...\}] \in S_n$$

We prove this by contradiction: a node $n$ is reachable by a source node $n_{so}$ through a direct predecessor $dp_i$. However $S_n$ lacks an entry which reflects this fact. Formally:

$$\exists n \in N, \exists dp_i \in DP_n, \exists P(n_{so}, dp_i) : \nexists [n_{so}, \{..., dp_i, ...\}] \in S_n$$

---

[3] Our proof technique was inspired by the one employed in [12].

This means that $dp_i$ did not include $n_{so}$ as an argument to the *sources* message (see Section 4.2.2). A node only sends the *sources* message if it has received the *sources* message from all its direct predecessors (see the sources Handler on line 10). Therefore, at least one of $dp_i$'s direct predecessors should have included $n_{so}$ as an argument to its *sources* message but failed to do so. Iteratively applying this reasoning would entail that $n_{so}$ did not include itself in the *sources* message to each of it direct successors. As shown in Algorithm 1 on line 8 this cannot be the case.

## F.2 Glitch Freedom

**Definition 5 (Glitch)** *Assume a node $n$ with update lambda $U_n$ and a set of direct predecessors $DP_n = \{dp \in N | dp >> n\}$. If $v_{dp_i}$ denotes a value propagated to $n$ by $dp_i$ and $Time_{n_{so}}(v_{dp_i})$ denotes a source node $n_{so}$'s timestamp attached to $v_{dp_i}$ then $U(v_{dp_i}, ..., v_{dp_{|DP_n|}})$ produces a glitch if:*

$$\exists P(n_{so}, dp_i) \land \exists P(n_{so}, dp_j) \land Time_{n_{so}}(val_{dp_i}) \neq Time_{n_{so}}(val_{dp_j})$$

We refer the reader to Section 4.1 for an intuitive explanation of glitches and glitch freedom.

### Theorem 2 (Glitch Freedom)

$$\forall n \in N : U_n(v_{dp_1}, ..., v_{dp_{|DP|}}) \iff \forall [n_{so}, \{dp_i, ..., dp_j\}] \in S_n : Time_{n_{so}}(v_{dp_i}) == Time_{n_{so}}(v_{dp_j})$$

Assume a set of source nodes $N_{so} = \{n_{so_1}, ..., n_{so_n}\}$ update and propagate their new values to all their direct successors concurrently. For a glitch to occur, at least one node $n$ in $PP_{N_{so}} = \bigcup_{i=1}^{n} PP_{n_{so_i}}$ needs to update itself with a set of values *vals* such that $\exists n_{so} \in N_{so}, \exists v_{dp_i}, v_{dp_j} \in vals : Time_{n_{so}}v_{dp_i} \neq Time_{n_{so}}v_{dp_j}$. A node is only able to invoke its update lambda if line 3 in the *change* Handler (see Section 4.2.4) returns a set of glitch free arguments. We discern two cases. First, $n$ was unable to find such a set of glitch free arguments. In this case $n$ could not have invoked its update lambda and could therefore not have caused a glitch which contradicts our assumption. Second, line 3 in the *change* Handler returned *vals* as set of glitch free arguments. In this case $n$ can safely invoke its update lambda using *vals* as arguments without causing a glitch which contradicts our assumption.

## F.3 Monotonicity

**Definition 6 (Monotonic update)** *Assume a node $n$ with update lambda $U_n$, a set of direct predecessors $DP_n = \{dp \in N | dp >> n\}$. $n$ invokes its update lambda using the following set of arguments $Args_t = \{v_{dp_1}, ..., v_{dp_{|DP_n|}}\}$. Moreover, $n$'s clock value is $t$ before this update happens. Later, at clock time $t + n$, $n$ invokes its update lambda using the following set of arguments: $Args_{t+n} = \{v'_{dp_1}, ..., v'_{dp_{|DP_n|}}\}$. We say that $n$ updates monotonically if and only if:*

$$\nexists dp_i \in DP_n, v_{dp_i} \in Args_t, v'_{dp_i} \in Args_{t+n} : Time_{n_{so}}(v_{dp_i}) > Time_{n_{so}}(v'_{dp_i}).$$

In other words, once $n$ updates with a value $v_{dp_i}$ originating from a source $n_{so}$ it will never update with an older value originating from the same source.

**Theorem 3 (Monotonicity)** *Nodes always update monotonically.*

We prove this by contradiction. Assume $n$ updates twice: first using $Args_t$ as set of arguments to $U_n$ and later using $Args_{t+n}$ as a set of arguments to $U_n$. We assume that $n$ updated non-monotonically, in other words the following holds:

$$\exists dp_i \in DP_n, v_{dp_i} \in Args_t, v'_{dp_i} \in Args_{t+n} : Time_{n_{so}}(v_{dp_i}) > Time_{n_{so}}(v'_{dp_i}).$$

Given that $v_{dp_i}$ and $v'_{dp_i}$ are the results of $dp_i$'s updates this means that $dp_i$ updated non-monotonically. Applying this reasoning iteratively results in a direct successor of $n_{so}$ updating non-monotonically. In other words, $n_{so}$ first propagated a value $v_{n_{so}}$ and then a value $v'_{n_{so}}$ for which the following holds:

$$Time_{n_{so}}(v_{n_{so}}) > Time_{n_{so}}(v'_{n_{so}})$$

However, this cannot happen given that clock times only monotonically increase (see line 6 in the *change* handler). Consequently, this means that none of $n_{so}$'s direct or indirect successors could have updated non-monotonically which contradicts our original assumption.

### F.4 Eventual Consistency

**Theorem 4 (DAG Construction)** *Any DAG can be constructed by recursively adding sink nodes (i.e. nodes with an out degree of 0) starting from the empty DAG.*

Every DAG has at least one topological ordering. Hence, one can construct any DAG by recursively adding nodes in the order for which they appear in the DAG's topological ordering. By the definition of topological ordering this entails that each node is added to the DAG before any of its successors. In other words, in each recursive step a node is added with out degree 0.

**Definition 7 (Consistency)** *A distributed dependency graph is consistent if the following holds:*

$$\forall n_{so} \in N, n \in PP_{n_{so}} : Time_{n_{so}}(lastProp(n)) = clock(n_{so}).$$

*We denote the value of source node $n_{so}$'s clock with $clock(n_{so})$ and we denote the last propagated value by $n$ with $lastProp(n)$.*

In other words, all nodes $n$ in a source node $n_{so}$'s propagation path must have witnessed its last update.

**Theorem 5 (Eventual Consistency)** *If all source nodes stop propagating new values, eventually the dependency graph reaches consistency.*

We prove this by structural induction over the dependency graph (i.e. a DAG constructed by recursively adding sink nodes starting from the empty DAG). The *induction base* considers a dependency graph containing a single node. This node is trivially consistent with itself. The *induction hypothesis* is that a given dependency graph $n$ reaches consistency when all of its source nodes stop propagating new values. The *induction step* extends this graph $n$ with a new *sink* node $n_{new}$. We do this by adding an arbitrary amount of edges from an arbitrary amount of nodes in $n$ to the new *sink* node. We prove by contradiction that this new graph is eventually consistent. Assume that all source nodes have stopped propagating updates and that our graph is inconsistent. Given our hypothesis this can only mean that $n_{new}$ causes the inconsistency. In other words:

$$\exists n_{so} \in N : n_{new} \in PP_{n_{so}} \wedge Time_{n_{so}}(lastProp(n_{new})) \neq clock(n_{so}).$$

There are two possible reasons for this:

First, $n_{new}$ was unable to update using its direct predecessors' last propagated value without causing a glitch. In other words, at least two of these values have a different *sClock* timestamp for $n_{so}$. However, the induction hypothesis ensures that $n_{new}$'s direct predecessors have witnessed $n_{so}$'s last update. Therefore, it is impossible for at least two of $n_{new}$'s direct predecessor to propagate values with a different *sClock* timestamp for $n_{so}$.

Second, $n_{new}$ was able to update itself using its direct predecessors' last propagated value, yet it still causes the graph's inconsistency. Through the *change* Handler (see Section 4.2.4) we know that $n_{new}$'s last propagated value's *sClocks* is a union of the *sClocks* of all values received from $n_{new}$'s direct predecessors. By definition this means that the following holds:

$$\forall dp \in \{dp \in N | dp >> n\} : Time_{n_{so}}(lastProp(dp)) = Time_{n_{so}}(lastProp(n_{new}))$$

Moreover, the induction hypothesis ensures that the following holds:

$$\forall dp \in \{dp \in N | dp >> n\} : Time_{n_{so}}(lastProp(dp)) = clock(n_{so}).$$

Therefore, $Time_{n_{so}}(lastProp(n_{new})) = clock(n_{so})$ which contradicts our original assumption.

### F.5 Progress

Informally, a distributed system makes progress if it performs useful computations towards termination [14]. In our case we define this termination as follows.

**Definition 8 (Update Completion)** *Assume a source node $n_{so}$ which updates and propagates a new value $v_{n_{so}}$. The update which caused $n_{so}$ to propagate $v_{n_{so}}$ completes if eventually:*

$$\forall n \in PP_{n_{so}} : Time_{n_{so}}(lastProp(n)) = Time_{n_{so}}(v_{n_{so}})$$

In other words, we say that a distributed reactive system provides progress if (concurrent) updates are guaranteed to finish in a finite amount of time. QPROP and QPROP$^d$ are therefore unable to guarantee progress given that both suffer from livelocks (see Section 7 and Appendix C.1 for examples and future avenues of research). In a nutshell, infinite concurrent updates to at least two source nodes can cause livelocks for common successors to said source nodes. However, assuming that concurrent updates stop, both algorithms are able to guarantee the completion of at least the last update to each source node. This trivially follows from the proof on eventual consistency (see Section F.4).

### F.6 Dynamic Graph Changes

Essentially, QPROP$^d$ provides two dynamic operations: adding and removing a dependency between nodes. Adding and removing a node from the dependency graph are sequences of dependency additions and removals. We therefore prove the correctness of dependency addition and removal.

**Dynamic Dependency Addition**    We refer the reader to Section 5.2 for a detailed explanation on the issue which can arise upon dynamically adding a dependency between two nodes in the dependency graph. Assume a node $n_1$ part of a source node's $n_{so}$'s propagation path $PP_{n_{so}}$. Moreover, assume $n_{so}$ propagates values $V_{before} = \{v_1, ..., v_n\}$ to its successors. A dependency is dynamically added from a node $n_2$ to $n_1$, which therefore adds $n_2$ to $PP_{n_{so}}$. After this, $n_{so}$ propagates values $V_{after} = \{v_{n+1}, ..., v_{n+n}\}$ to its successors However, $n_2$ will never receive values from $V_{before}$ given that these were propagated before it joined $n_{so}$'s propagation path. QPROP$^d$ must therefore ensure that all nodes in $PP_{n_{so}}$ update using values from $V_{before}$ before updating using values from $V_{after}$.

We prove this by contradiction. Assume a node $n_3$ in $PP_{n_{so}}$ updates using values from $V_{after}$ without first updating using values from $V_{before}$. Given that nodes propagate values in order, this can only mean that $\exists P(n_2, n_3)$ (i.e. all other nodes in $PP_{n_{so}}$ will first propagate values from $V_{before}$). According to Algorithm 3 $n_3$ must have received the *addSources* message from at least one of its predecessors $n_{3_{pred}}$. Moreover, $n_{3_{pred}}$ must be a direct successor of $n_2$ and must have added $n_2$ to its $B_r$ dictionary. Line 11 in Algorithm 4 ensures that $n_{3_{pred}}$ will only update itself using the first value of $V_{after}$ (i.e. $v_{n+1}$) if it previously updated itself with the last value of $V_{before}$ (i.e. $v_n$). Hence, $n_3$ can impossibly receive values from $V_{after}$ before receiving values from $V_{before}$.

**Dynamic Dependency Removal**    We refer the reader to Section D.2 for a detailed explanation on the issue which can arise upon dynamically removing a dependency between two nodes in the dependency graph. Dynamically removing a dependency between a node $n_2$ and its predecessor $n_1$ changes $n_2$'s $S$ dictionary as well as the $S$ dictionaries of all its successors. QPROP$^d$ must therefore ensure glitch freedom while nodes update their topological information.

A node $n$ only updates its topological information as a result of receiving the *remSources* message (see remSources Handler). We discern two cases. First, $n$ removes an entry for a source node $n_{so}$ from $S$ (see line 5 in the remSources Handler). In this case, $n$ only had a single predecessor propagating values which originate from $n_{so}$ and can therefore not produces glitches. Second, $n$ removes a predecessor *pred* from an entry for a source node $n_{so}$ which still contains other predecessors *preds*. However, in this case $n$ removes all values for *pred* from $I$. All subsequent values $n$ will receive from *pred* will no longer originate from $n_{so}$ and can therefore not cause glitches.

## About the authors

**Florian Myter** is a PhD student at the Software Languages Lab, Vrije Universiteit Brussel in Belgium. His main research area is distributed programming and more concretely the design and implementation of programming techniques to deal with distributed state. Contact him at fmyter@vub.be 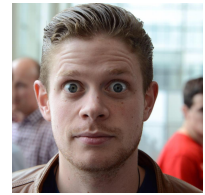

**Christophe Scholliers** is professor in foundations of programming languages at Ghent University. His current research is mainly situated in the field of parallel and distributed programming language abstractions. Contact him at christophe.scholliers@ugent.be 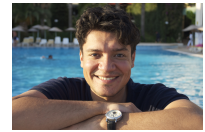

**Wolfgang De Meuter** is a professor in programming languages and programming tools. His current research is mainly situated in the field of distributed programming, concurrent programming, reactive programming and big data processing. His research methodology varies from more theoretical approaches (e.g. type systems) to building practical frameworks and tools (such as e.g. crowdsourcing systems). Contact him at wdmeuter@vub.ac.be 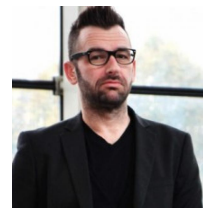